\documentclass[]{aastex631}

\newcommand{\mc}{\multicolumn}

\begin{document}

\title{COSMIC: A Galaxy Cluster Finding Algorithm Using Machine Learning}

\correspondingauthor{Jun-Qing Xia, Zhong-Lue Wen}
\email{xiajq@bnu.edu.cn, zhonglue@nao.cas.cn}

\author{Da-Chuan Tian}
\affiliation{Institute for Frontiers in Astronomy and Astrophysics, Beijing Normal University, Beijing 100875, China}
\affiliation{School of Physics and Astronomy, Beijing Normal University, Beijing 100875, China}

\author{Yang Yang}
\affiliation{Beijing Navigation School, Beijing 101118, China}

\author{Zhong-Lue Wen}
\affiliation{National Astronomical Observatories, Chinese Academy of Sciences, 20A Datun Road, Chaoyang District, Beijing 100101, China}

\author{Jun-Qing Xia}
\affiliation{Institute for Frontiers in Astronomy and Astrophysics, Beijing Normal University, Beijing 100875, China}
\affiliation{School of Physics and Astronomy, Beijing Normal University, Beijing 100875, China}



\begin{abstract}

Building a comprehensive catalog of galaxy clusters is a fundamental task for the studies on the  structure formation and galaxy evolution. In this paper, we present COSMIC (Cluster Optical Search using Machine Intelligence in Catalogs), an algorithm utilizing machine learning techniques to efficiently detect galaxy clusters. COSMIC involves two steps, including the identification of the brightest cluster galaxies and the estimation of the cluster richness. We train our models on the galaxy data from the Sloan Digital Sky Survey and WHL galaxy cluster catalog. Validated to a test data in the region of northern Galactic cap, COSMIC algorithm demonstrates a high completeness when cross-matching with previous cluster catalogs. Richness comparison with previous optical and X-ray measurements also demonstrated a tight correlation. Our methodology showcases robust performance in galaxy cluster detection and holds promising prospects for applications in upcoming large-scale surveys.

\end{abstract}

\keywords{Galaxy clusters (584), Brightest cluster galaxies (181), Classification (1907), Convolutional neural networks(1938)}


\section{Introduction} \label{sec:intro}

As the most massive gravitationally bound systems in the universe, galaxy clusters are one of powerful cosmological probes \citep{Majumdar2004, Hu2003, Lima2004, Lima2005}. Their abundance and spatial distribution provide the ways to verify the growth of cosmic structure and constrain cosmological parameters \citep{evrard_biased_1989, 1992A&A...262L..21O,allen_cosmological_2011}. 
The high density makes galaxy clusters act as idea laboratories to reveal the nature of dark matter and evolution of galaxies. Observations in optical, X-ray and radio wavelengths show intriguing features from  different components in galaxy clusters. Therefore, galaxy clusters are one of the most important targets in the past or ongoing survey projects, e.g. Sloan Digital Sky Survey \citep{york_sloan_2000}, 
Chandra \citep{weisskopf2000chandra}, XMM-Newton \citep{jansen2001xmm} and Dark Energy Survey \citep{Abbott2018}, and in the future projects, e.g. Euclid satellite \citep{Laureijs2011}, the Vera C. Rubin Observatory \citep{2019ApJ...873..111I} and Chinese Space Station Telescope \citep{zhanhu11}.

The multiple components of galaxy clusters emit radiations that make them visible in optical, X-ray and radio wavelengths. Optical observations detect the discrete member galaxies concentrated in clusters. Spectroscopic surveys are the most powerful way to identify galaxy clusters in 3 dimensional space, but currently only for nearby galaxy clusters. Optical photometric data provide the most number of galaxy clusters up to high redshifts \citep{Wen2018, Wen2022}. In X-ray, galaxy clusters have an extended morphology from hot intracluster gas, by which X-ray clusters can be efficiently found from the  imaging surveys, e.g. ROSAT \citep{Truemper1982, Voges1999} and eROSITA \citep{Merloni2012, Predehl2021}. The hot gas in clusters makes distortion on the spectrum of the cosmic microwave background, known as Sunyaev–Zeldovich (SZ; \citealt{1972CoASP...4..173S}) effect. The brightness of the SZ signal is independent of redshift, so that clusters can be detected up to high redshifts via SZ effect from the surveys, e.g. Planck \citep{Tauber2004}, SPT \citep{Bleem2015} and ACT \citep{Hilton2021}.
While various methods have been developed, observations in optical play an essential role for building a more complete catalog of galaxy clusters benefited by many large imaging surveys. In light of the unprecedented galaxy data presently available and in the future, the establishment of an increasingly comprehensive galaxy cluster catalog becomes achievable, thereby presenting challenges to our existing methodologies for galaxy cluster detection.

In optical, galaxy clusters show a distinct concentration of galaxies.
Based on multi-band optical survey, e.g. SDSS, many methods have been presented to detect galaxy clusters by searching for overdensity of galaxies on the sky with background galaxies statistically separated. The galaxies in a cluster show a tight color--magnitude relation, known as red sequence, which is an efficient way to separate member galaxies and background galaxies. The red-sequence based algorithms include maxBCG \citep{koester_maxbcg_2007}, GMBCG \citep{hao_gmbcg_2010}, CAMIRA \citep{Oguri2014} and redMaPPer \citep{Rykoff2016}. Another kind of method directly uses photometric redshifts of galaxies to
separate member galaxies and background galaxies. The photometric redshift based algorithms include
WHL \citep{Wen_2009, wen_catalog_2012, wen_calibration_2015}, AMF \citep{szabo_optical_2011}. Some of the algorithms also include the features of the brightest cluster galaxy (BCG), which are the most massive galaxies and exhibit a distinctive red color relative to other member galaxies in the clusters. Typically, inclusion of the BCGs helps to identify galaxy clusters more efficiently and determine the cluster centers. The above algorithms often focus on identifying galaxy clusters with specific characteristics, such as the red sequence, density profile and luminosity function. As a result, achieving high completeness becomes challenging for clusters lacking such distinctive features, especially for clusters at high redshifts or with a low mass. To create a more comprehensive catalog of galaxy clusters, it is imperative to develop new techniques that can effectively leverage both existing and future optical survey data.

Artificial intelligence techniques, such as machine learning and deep learning, have the advantage of being able to figure out complex non-linear relationships between parameters.
They offer a diverse array of algorithms, ranging from source classification \citep{Fadely2012,Kim_2016,hughes_quasar_2022} to parameter estimation \citep{Hezaveh2017, huang_strong_2022}. They typically yield results comparable to, or even surpassing, those obtained by traditional approaches. Recent applications of machine learning for galaxy cluster detection in optical  include studies by \citet{grishin_yolo-cl_2023} and \citet{chan_deep-cee_2019}, where two types of object detection algorithms are applied to detect galaxy clusters in color images from the SDSS. Similar to real-time target detection in practical scenarios, the identified target in this context will be outlined by a bounding box, along with a predicted probability indicating the likelihood of it being classified as a galaxy cluster. According to \citet{grishin_yolo-cl_2023}, these methods aim to mitigate potential systematic uncertainties introduced during source detection, photometry, and photometric measurements. Furthermore, they achieve a higher level of cluster catalog completeness and purity compared to traditional cluster detection algorithms typically employed in SDSS. Therefore, machine learning techniques are expected to have great potential for searching galaxy clusters.

In this work, we present COSMIC (Cluster Optical Search using Machine Intelligence in Catalogs), an algorithm utilizing machine learning techniques to efficiently detect galaxy clusters.
COSMIC algorithm consists of two steps.
First, we utilize an XGBoost classifier to identify BCG-like galaxies from photometric and spectroscopic data.
Then, we employ a trained deep neural network to estimate cluster richness as a proxy for cluster mass from the smoothed optical map obtained around each BCG-like galaxy.
This study exclusively utilizes the SDSS data to ensure its effectiveness and applicability in current and forthcoming sky surveys. The paper is structured as follows. In Section \ref{sec:data}, we describe the galaxy and galaxy cluster data required for training machine learning models. Section \ref{sec:methodology} is devoted to the model design and training. Our results are presented in Section \ref{sec:resultsANDdiscussion} where we evaluate the performance of COSMIC algorithm within a test area of $\sim$ 200 square degrees in the sky. Finally, we summarise our algorithm and results in Section \ref{sec:summaryANDperspective}, and highlight potential applications of galaxy surveys. Throughout this work, we assume a $\Lambda$CDM cosmology, taking $H_0 = 100h$ km s$^{-1}$ Mpc$^{-1}$, with $h = 0.7$, $\Omega_m = 0.3$ and $\Omega_\Lambda = 0.7$.

\section{Data preparation for training} \label{sec:data}
We use the data of galaxies and galaxy clusters in the SDSS to develope COSMIC algorithm.
Here, we describe the training samples and data pre-procession suitable for our machine learning models, named BCG Classifier and Richness Estimator.

\subsection{WHL clusters and member galaxy candidates} \label{sec:WHL clusters and member galaxy candidates}

Based on photometric redshifts of galaxies, \citet{wen_catalog_2012} identified 132,684 galaxy clusters in the redshift of $0.05 < z < 0.8$. They assumed that each luminous galaxy is located 
in a cluster candidate. The member galaxy candidates of the cluster candidate are obtained within
a photometric redshift slice and a projection separation. The redshift of the cluster candidate 
is being the median value of photometric redshifts of member candidates. The position of the brightest galaxy is 
taken as the center of the cluster candidate. The richness of the cluster candidate is estimated
based on the total luminosity of member candidates. Finally, a cluster is identified with a 
richness threshold.
An updated version of WHL catalog is published with a new richness estimate as the proxy of cluster mass and updated spectroscopic redshifts of clusters. 
Plus the newly identified 25419 clusters at high redshifts, the new WHL catalog contains 158,103 clusters in total~\citep{wen_calibration_2015}. 
The spectroscopic redshifts of clusters are determined 
from spectroscopic redshift available member galaxies, most of which are BCGs.
About 89\% of clusters have spectroscopic redshifts in the SDSS spectroscopic survey region (covering the northern and southern Galactic caps). 
Most clusters have a richness of less than 200 and the number of member galaxies $N_{500} \ge 6$ within a radius of $r_{500}$, where $r_{500}$ is the radius within which the mean density of a cluster is 500 times the critical density of the universe. 
The cluster richness is defined to be the $r$-band total luminosity of member galaxies within $r_{500}$ which can act as a a well-calibrated mass proxy in optical \citep{wen_calibration_2015}. 
This galaxy cluster catalog has the largest number of clusters and high completeness among the cluster catalogs based on the SDSS data, and a wide range of redshift, as well as the positions of their BCGs. 
Therefore, the WHL cluster catalog is a good sample for training BCG Classifier and Richness Estimatior. 

To have reliable photometric data, we merely use the SDSS data in the high Galactic latitude region with spectroscopic observations. We get member galaxies of WHL clusters within a projection radius of 1 Mpc 
from the SDSS data. 
If the galaxies have spectroscopic redshifts, they are regarded as member galaxies if they have a velocity difference $\Delta v < 2500$ km~s$^{-1}$ from the cluster redshifts. 
Otherwise member galaxies are selected whinin a
photometric redshift slice according to the uncertainty of photometric
redshift \citep{wen_calibration_2015}. The half thickness of the photometric redshift slice is in the form of
\begin{equation}
  \Delta z=\left\{
\begin{array}{ll}
   0.04\,(1+z)           &     \mbox{for $z\leq0.45$}\\
   0.248\,z-0.0536       &     \mbox{for $z>0.45$}
\end{array}
\right. \,,
\end{equation}
The galaxies are limited to have an evolution-corrected absolute magnitude of $M_e\le-20.5$, where $\mathrm{M_r^e}$ is the evolution-corrected absolute magnitude in $r$-band, $\mathrm{M_r^e}(z) = \mathrm{M_r}(z) + 1.16z$~\citep{wen_calibration_2015}. The so-selected
member galaxies have a completeness about 90\% and a
contamination rate about 20\% within the given photometric
redshift slice \citep{Wen_2009}

\subsection{Dataset for training BCGs} \label{sec:Dataset for training BCGs}

We get photometric and spectroscopic data of galaxies from SDSS SkyServer DR17\footnote{\href{https://skyserver.sdss.org/dr17/SearchTools/sql}{https://skyserver.sdss.org/dr17/SearchTools/sql}}~\citep{Abdurro’uf2022}, in which the misclassified stars and QSOs with spectral are removed. 
Then, this dataset is cross-matched with the BCGs of the WHL galaxy clusters, resulting in corresponding BCG labels along with corresponding WHL galaxy cluster information such as richness and the number of member galaxies. 
This catalog is used to train our BCG Classifier containing overall 210780 galaxies (hereafter referred to as PhotCat) which includes matched WHL BCGs with a richness greater than 10 as positive samples and an equal number of non-BCGs as negative samples randomly selected. We ensure that the ratio of positive to negative samples is approximately 1 to mitigate the impact of sample imbalance for classification.

\begin{figure}
    \includegraphics[width=\columnwidth]{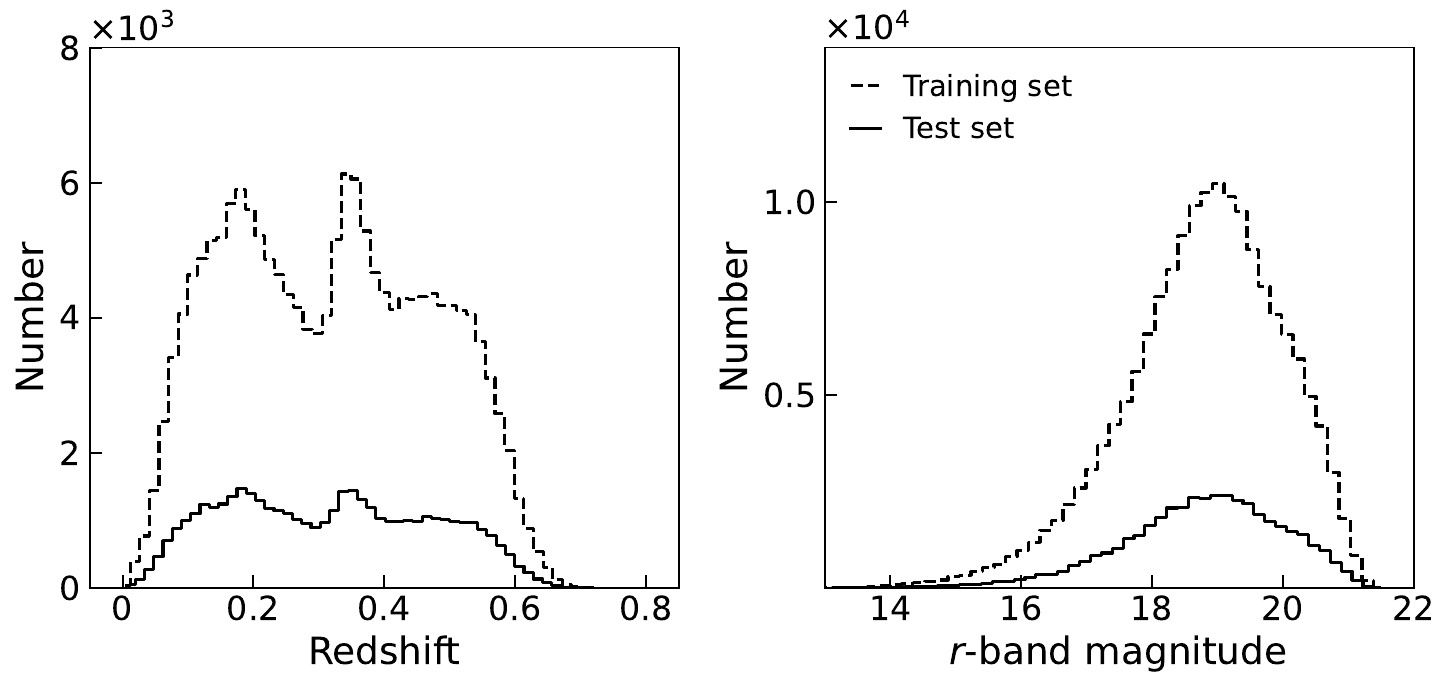}
    \caption{The redshift and $r$-band magnitude distribution of galaxies of PhotCat in the training set (dashed lines) and test set (solid lines).}
    \label{fig:figure1}
\end{figure}

Same as the most common way to train a machine learning model, PhotCat are randomly divided into two sets with 80 percent as the training set and the rest as the test set, where the former is used to train our model, i.e. adjust various parameters of the model, and the latter is used to evaluate the performance of the model. The redshift and $r$-band magnitude distributions of both the training and test sets are illustrated in Figure \ref{fig:figure1}. Notably, galaxies are distributed across various redshift intervals, primarily within the range of $0.01 < z < 0.80$, which is indicative of a representative sample of the overall galaxy population. In the figure, the dashed lines represent the training set, while the solid lines represent the test set. 

\subsection{Smoothed optical map for training cluster richness} \label{sec:Smoothed optical map for training cluster richness}

\begin{figure}
    \includegraphics[width=\columnwidth]{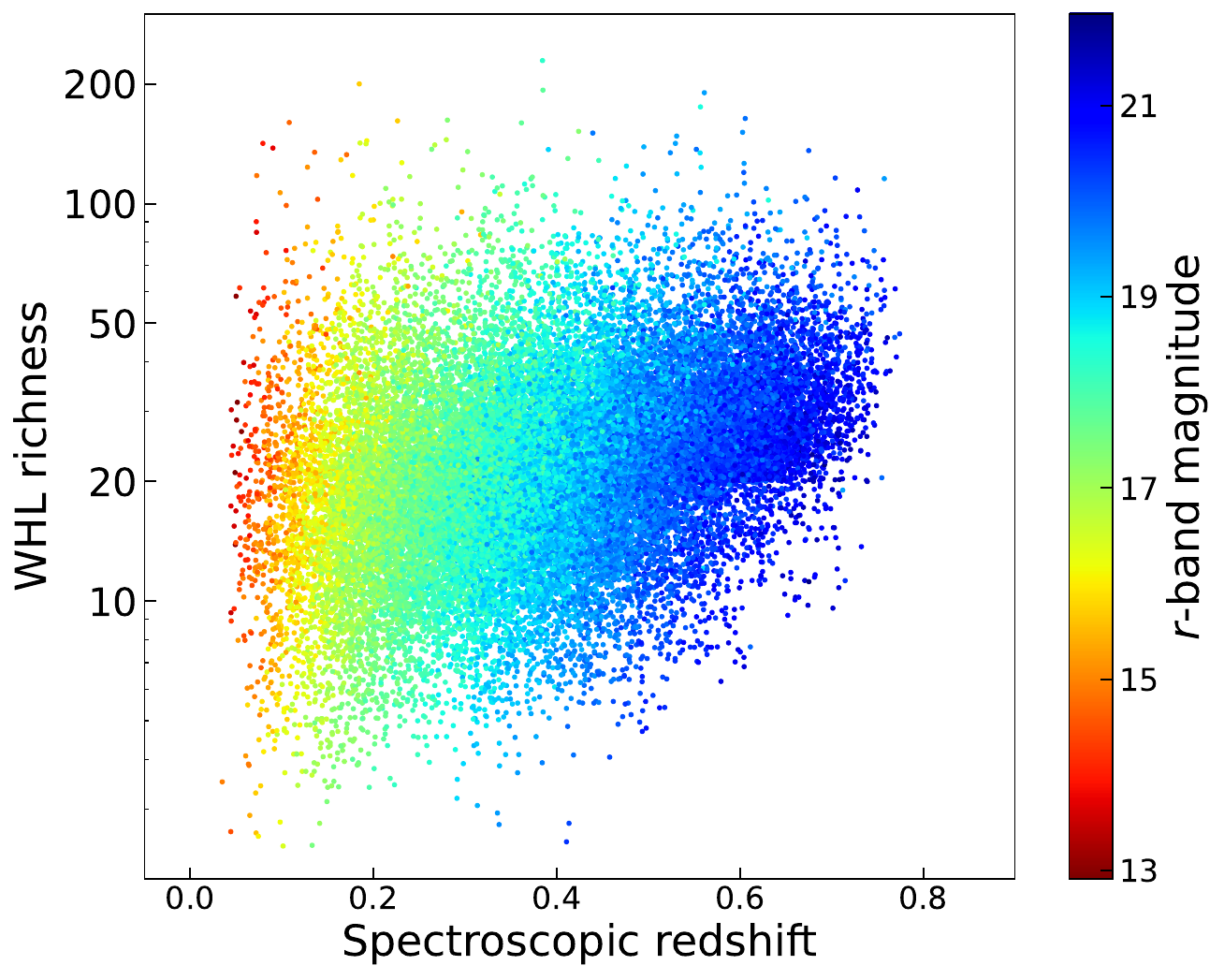}
    \caption{The richness distribution of WHL clusters with spectroscopic redshifts. The colorbar represents the $r$-band magnitude of their BCGs. Only 20\% of the WHL clusters are randomly selected for plotting.}
    \label{fig:figure2}
\end{figure}

\citet{wen_calibration_2015} calibrated the cluster richness as an optical mass proxy of galaxy clusters. We use this metric to evaluate whether there are galaxy clusters around the BCG candidates selected in the first step. Given the critical significance of accurate redshifts, galaxy clusters in the WHL catalog with spectroscopic redshifts are used to train our Richness Estimator. Their distributions in redshift, richness and BCG $r$-band magnitude are shown in Figure \ref{fig:figure2}.

The relationship between the richness and the total luminosity inspire us to utilize information extracted from optical images to estimate galaxy cluster richness. To this end, we make the smoothed optical maps (SOM) of clusters similar as~\citet{Wen2013} by mapping $r$-band luminosity information of galaxies surrounding the BCG within 1 Mpc of each cluster to a 2-dimensional map and then applying Gaussian Smoothing. All SOMs within a $2\times 2$ Mpc$^2$ physical size centered on BCGs of each clusters are resized to 200 pixel $\times$ 200 pixel. Details of producing these SOMs are as follows:

\begin{itemize}
    \item[1.] We transfer the location of member galaxies from the equatorial coordinates (R.A., Dec.) to Cartesian coordinates $(x,y)$ centered on the BCG. A physical size of the sky area of 2 Mpc $\times$ 2 Mpc is transferred to an image matrix with a dimension of 200 pixel $\times$ 200 pixel. 
    \item[2.] The luminosity corresponding to each pixel is obtained by convolving the luminosity of all member galaxies with a Gaussian kernel, i.e. 
        \begin{equation}
        I(x_i,x_j)=\sum_{k=1}^{N} L_k g(x_i - x_k, y_j - y_k, \sigma_k),
        \label{eq:lum}
        \end{equation}
    where $x_k$ and $y_k$ are coordinates of the $k$th member galaxy with its $r$-band luminosity $L_k$ in units of $L^\ast$, $N$ is the number of member galaxies. And the 2-dimensional Gaussian function $g(x,y,\sigma)$ with a smoothing scale $\sigma=$ 0.05 $\times$ 1 Mpc is
        \begin{equation}
        g(x,y,\sigma) = \frac{1}{2\pi \sigma^2} \exp \left( -\frac{x^2 + y^2}{2\sigma^2} \right).
        \label{eq:gauss}
        \end{equation}
    \item[3.] Produce SOM with the luminosity matrix obtained from procedures above.
\end{itemize}

To subtract the background of each cluster, we use a method similar to that described in \citet{wen_calibration_2015}. The local background is estimated by dividing the region between 2 and 4 Mpc around each galaxy cluster into 48 sectors with an equal area. Field galaxies fainter than the second brightest cluster member are used to calculate the background. The sectors with luminosities significantly deviating from the mean by 3\,$\sigma$ are excluded, and the background is recalculated and then subtracted from each pixel of the luminosity matrix.
However, the distance information along the line of sight are not taken into account of SOMs, which could result in confusions that the distant clusters seem to be dimmer than those closer to us for clusters with the similar richness. Thus, a redshift correct is adopted to this bias, i.e. multiply each image matrix by E(z)
   \begin{equation}
   E(z) = \sqrt{\Omega_\Lambda + \Omega_m(1+z)^3}.
   \label{eq:Ez}
   \end{equation}
    
\begin{figure*}
    \includegraphics[width=\textwidth]{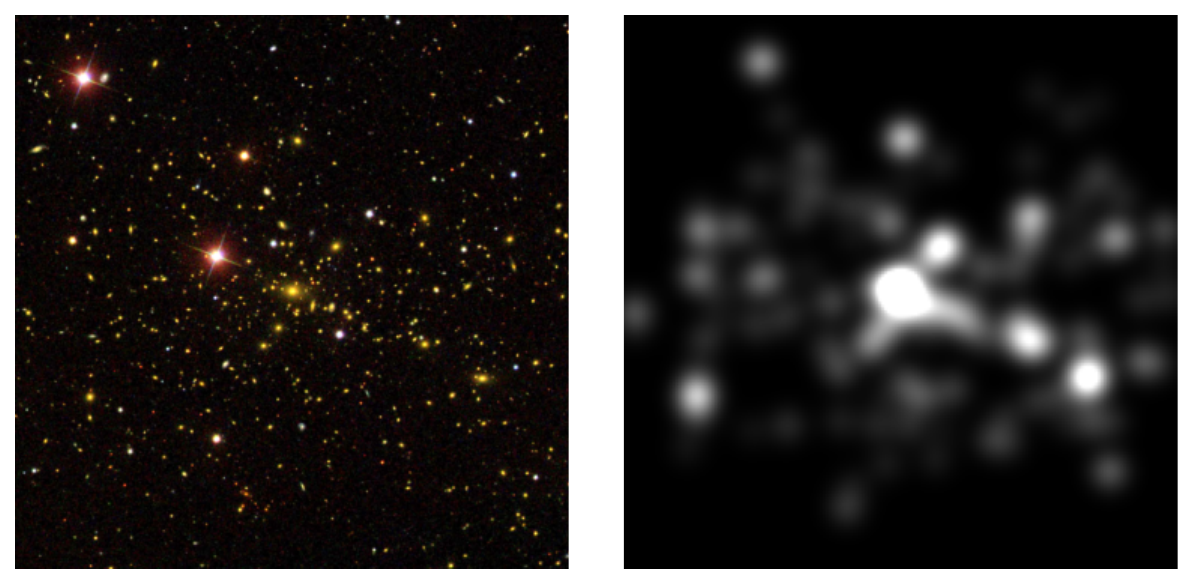}
    \caption{An example of smoothed optical map (right panel) of a WHL cluster with its corresponding color image (left panel). The size of SOM is 200 pixel $\times$ 200 pixel over a physical area of $2\times 2$ Mpc$^2$ and the physical scale of color image is same as the SOM.}
    \label{fig:figure3}
\end{figure*}

An example is shown in Figure \ref{fig:figure3}. Here is an SOM example of a WHL cluster in the right panel with its corresponding color image from SDSS DR16 Image List Tool\footnote{\href{http://skyserver.sdss.org/dr16/en/tools/chart/listinfo.aspx}{http://skyserver.sdss.org/dr16/en/tools/chart/listinfo.aspx}} in the left panel. 
It should be noted that SOMs are not normalized, and as a result, fainter member galaxies may not be as easily visible compared to the brighter galaxies. Thus, for a better display, we have appropriately amplified the brightness of the SOM example.
SOMs provide position and photometric information for galaxies near the center of galaxy clusters, excluding interference from some field galaxies. They can serve as a rough proxy for optical cluster images.

\begin{figure}
    \includegraphics[width=\columnwidth]{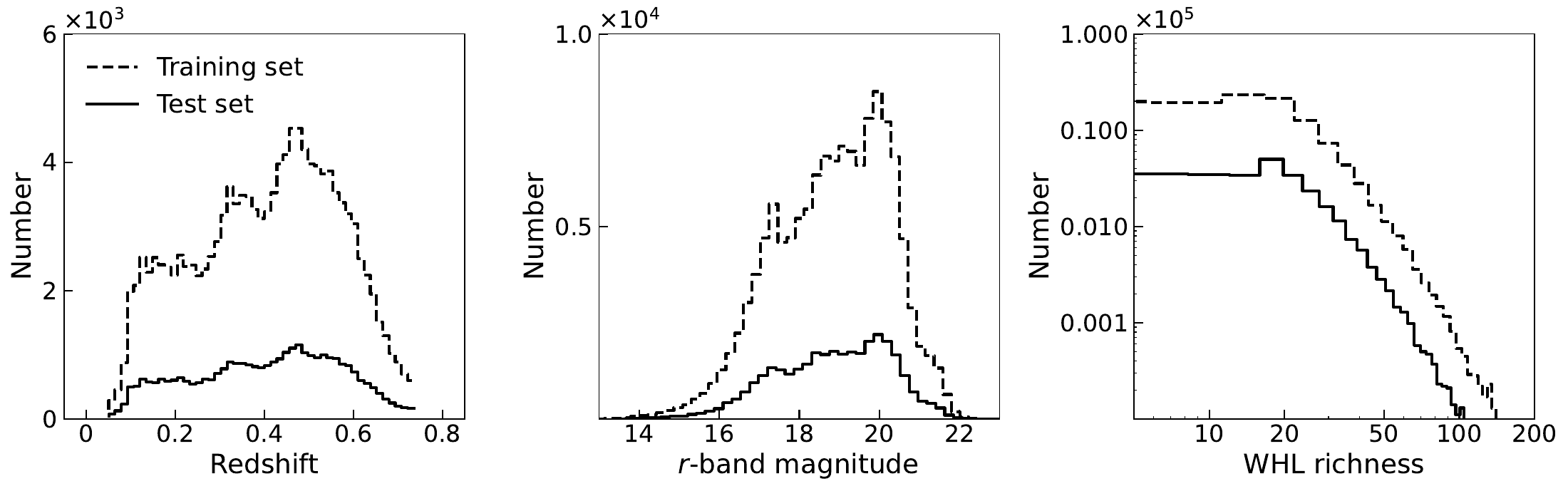}
    \caption{The spectroscopic redshift, $r$-band magnitude and richness distribution of samples in the training set (dashed lines) and the test set (solid lines) used for training Richness Estimator. The left two panels are for BCGs in the positive samples and for normal galaxies in the training data.}
    \label{fig:figure4}
\end{figure}

In this study, a total of 163,021 SOMs are utilized to train and assess our Richness Estimator. This training data comprises SOMs of both WHL galaxy clusters with spectroscopic redshifts (positive samples) and normal field galaxies (negative samples), where the negative samples source from spectral galaxies with R.A. from 150 to 180 degree and Dec. from 44 to 54 degree. We assess the richness of the negative samples using the same algorithm applied to WHL clusters and this part of the training set allows the model to make more reasonable richness estimations. Also, the training data are randomly divided into two sets: 80 percent are used for training the Richness Estimator, while the remaining 20 percent serve as the test set to evaluate its performance.
The distributions of both training set and test set are shown in Figure \ref{fig:figure4}. The left two panels present the distributions of spectroscopic redshifts and $r$-band magnitudes for BCGs in the positive sample and for normal field galaxies in the both sets. As we can see, most samples have a redshift range of $0.05 < z < 0.75$. 

\section{Methodology} \label{sec:methodology}

\begin{figure}
    \includegraphics[width=\columnwidth]{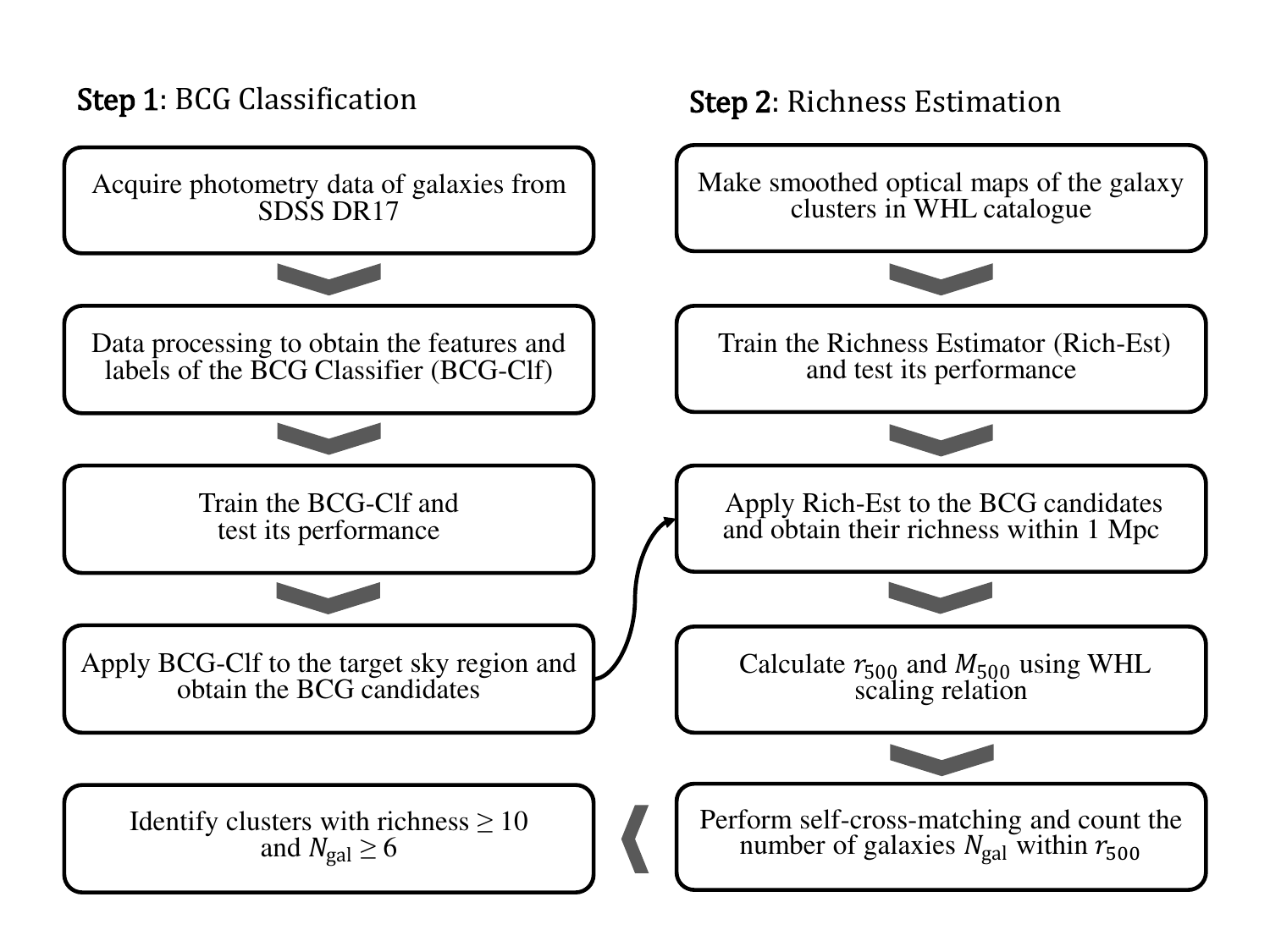}
    \caption{Full workflow. COSMIC algorithm consists of two main steps, BCG Classification and Richness Estimation.}
    \label{fig:figure5}
\end{figure}

Figure \ref{fig:figure5} outlines COSMIC algorithm to detect galaxy clusters. In brief, there are two main steps, including BCG Classification and Richness Estimation where two machine learning techniques are adopted, XGBoost and deep neural network, respectively. Below we provide detailed descriptions of the model architectures and training procedures.

\subsection{Step 1: BCG Classification} \label{sec:BCG Classification}

The BCGs are the most massive galaxies in the Universe. They are very luminous in optical and have a red color. In the color--redshift and magnitude--redshift diagrams, they are located in the red end and bright
end, which are distinguished from other galaxies. In morphology, the BCGs show an extended envelope compared to
the normal elliptical galaxies. Here, we present a model named BCG Classifier based on XGBoost trained by PhotCat.

\subsubsection{XGBoost and performance metrics of classification task} \label{sec:xgboost}

XGBoost, short for eXtreme Gradient Boosting~\citep{chen_xgboost_2016}, is a powerful machine learning technique widely used for regression and classification tasks. It represents one of the fastest implementations of gradient-boosted decision trees. In this method, decision trees are employed as weak learners, and their estimates are weighted and combined to generate the final predictions. By aggregating the predictions of multiple weak learners, known as strong learners, the XGBoost achieves significantly improved performance compared to individual decision trees. The XGBoost has gained popularity due to its ability to handle complex datasets, its efficiency in training models, and its robustness against overfitting. It provides advanced regularization techniques, such as shrinkage and pruning, to enhance model generalization. The method also incorporates optimization algorithms that efficiently minimize the loss function, resulting in faster convergence and improved prediction accuracy. For a comprehensive understanding of the XGBoost and its various features, the reader is recommended referring to the official online documentation.\footnote{\href{https://xgboost.readthedocs.io/}{https://xgboost.readthedocs.io/}}

In our study, we train a BCG classifier based on the XGBoost classifier. Our input features consist of a set of observational data, including the $r$-band magnitude, the color $u-r$, $g-i$, $r-i$, and $i-z$, as well as $\mathrm{deVRad^r}$ and $\mathrm{deVAB^r}$, along with their respective errors, and the photometric redshift.
Due to the inherent characteristics of decision trees, it is unnecessary to implement a normalization strategy for these features. With the above features of the galaxy, the model can provide the probability of the galaxy being a BCG.

We employ a 5-fold grid search cross-validation approach \citep{pedregosa2011scikit} to determine the optimal hyperparameters for our model. The cross-validation function provided by the XGBoost package facilitates this process by implementing grid search. 
For comprehensively evaluating the model performance, we utilize the Area Under the Curve (AUC) metric during the training process. The AUC is computed as the area under the receiver operating characteristic (ROC) curve, which is a line connected by the true positive rate (TPR) and false positive rate (FPR) under different thresholds from 0 to 1. The TPR is calculated as the ratio of true positives (TP) to the sum of TP and false negatives (FN), where true positives are cases correctly predicted as positive and false negatives are cases incorrectly predicted as negative. The FPR is calculated as the ratio of false positives (FP) to the sum of FP and true negatives (TN), where false positives are cases incorrectly predicted as positive and true negatives are cases correctly predicted as negative. The definitions of the TPR and FPR are also provided in Eq. \ref{eq:TPRandFPR}.
The threshold represents the probability that a classifier determines a sample as a positive. The area between the ROC curve and the axes is the AUC. Thus, the AUC reflects the classification ability of the classifier for both positives and negatives. A perfect model has an AUC of 1. The AUC aims to train a model that minimizes false positives, aligning with our goal in the BCG classification stage.
\begin{equation}
    \mathrm{TPR} = \frac{\mathrm{TP}}{\mathrm{TP}+\mathrm{FN}},\quad \mathrm{FPR} = \frac{\mathrm{FP}}{\mathrm{FP}+\mathrm{TN}}.
    \label{eq:TPRandFPR}
\end{equation}

\begin{equation}
    \mathrm{Accuracy} = \frac{\mathrm{TP}+\mathrm{TN}}{\mathrm{TP}+\mathrm{TN}+\mathrm{FP}+\mathrm{FN}},\quad \mathrm{Recall} = \frac{\mathrm{TP}}{\mathrm{TP}+\mathrm{FN}}.
    \label{eq:AccuracyandRecall}
\end{equation}

\begin{figure}
    \includegraphics[width=\columnwidth]{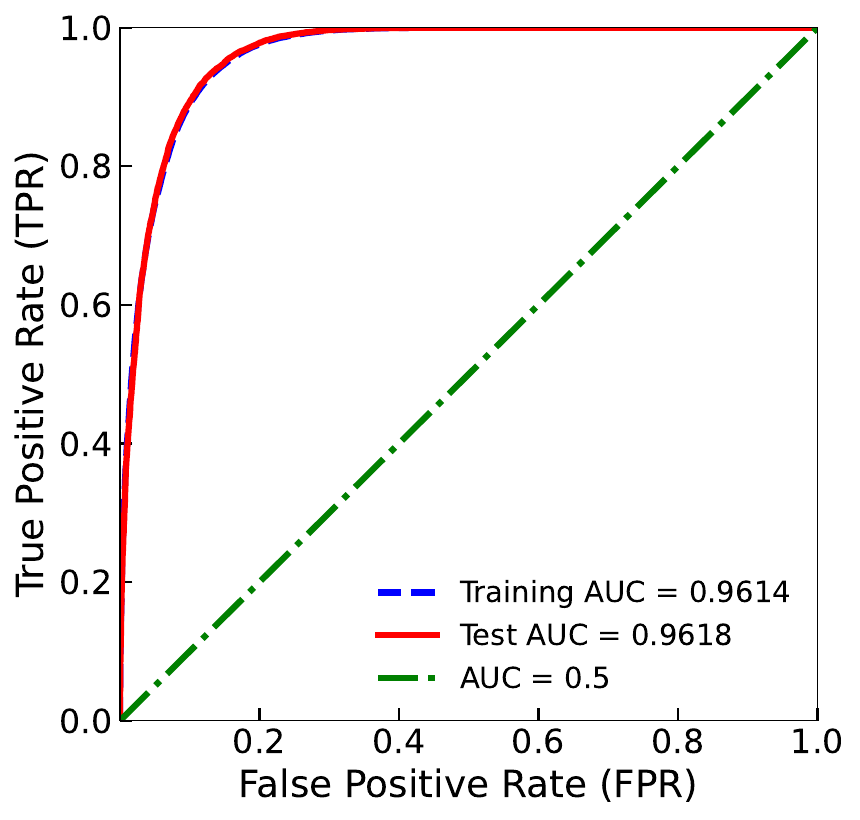}
    \caption{The ROC curves of training set and test set. The AUC value is the area between the curve and the axis. The green dashed line represents an AUC value of 0.5, which means the model has no ability of classification.}
    \label{fig:figure6}
\end{figure}

With the finalized hyperparameters and the training data, we construct a BCG classifier reaching an accuracy, recall and AUC of 90.0\%, 93.4\% and 0.9614 for the training set, and 90.2\%, 93.4\% 
and 0.9618 for the test set, respectively. 
The accuracy and recall we used here are defined as Eq. \ref{eq:AccuracyandRecall}.
Figure \ref{fig:figure6} shows the ROC curves and AUC of the training set and test set. These metrics collectively demonstrate the robust performance of our model in classifying BCG and non-BCG.

Unlike traditional methods that rely on color magnitude diagrams to differentiate between BCGs and ordinary galaxies, XGBoost operates in multiple dimensions, enabling it to make decisions based on a wider range of features and parameters. This multidimensional approach enhances the discriminatory power of the model, allowing for more accurate and precise classification of BCGs. By leveraging the capabilities of XGBoost, we can exploit its advantage over traditional methods and achieve improved performance in BCG classification tasks.

\subsection{Step 2: Richness Estimation} \label{sec:Smoothed optical map for training cluster richness}

Richness is usually regarded as a proxy of cluster mass. 
Here, we present the second stage of COSMIC algorithm, referred to as Richness Estimation. In this stage, we focus on estimating the richness of galaxy clusters centered on the BCGs based on the smoothed optical maps.

To estimate cluster richness from SOMs, we employ a transfer learning strategy, which involves utilizing a pre-trained deep neural network and fine-tuning it according to the task at hand. The neural network we used is a residual neural network (ResNet;~\citealt{He2015}). Compared to traditional neural networks, such as CNNs, the ResNet achieves higher accuracy and maintains a faster converging rate by introducing the residual learning module with shortcut connections. A detailed description of the building blocks and connection skipping can be found in~\citet{He2015}. The reference proposed five kinds of ResNet models, going up to 152 layers. Considering the limited amount of data, computing resources, and model performance comparison, we select the ResNet-34 pre-trained model, which consists of 34 layers for feature extraction from images. The ResNet-34 module is implemented in torchvision.models\footnote{\href{https://pytorch.org/vision/stable/models.html}{https://pytorch.org/vision/stable/models.html}} and obtained from the PyTorch deep learning library. By employing the ResNet-34 architecture, we aim to harness its depth and feature extraction capabilities to achieve accurate richness estimation for galaxy clusters from SOMs.

The original ResNet model is designed for classification tasks, so it needs to be modified to be suitable for richness estimation. We employ fully-connected (FC) layers to replace the original classification network of ResNet. The FC layer means that all input units from the last layer are connected to all units in the next layer, which is the most common layer in deep learning. The purpose of the FC layer is to compile the global information extracted from the previous layers to form the final output, which can be written as Eq. \ref{eq:FClayer}:
\begin{equation}
    \mathrm{FC}(x) = W^{T} x + b
    \label{eq:FClayer}
\end{equation}
where $W$ and $b$ represent weights and biases, respectively, are learnable parameters. 

\begin{figure}
    \includegraphics[width=\columnwidth]{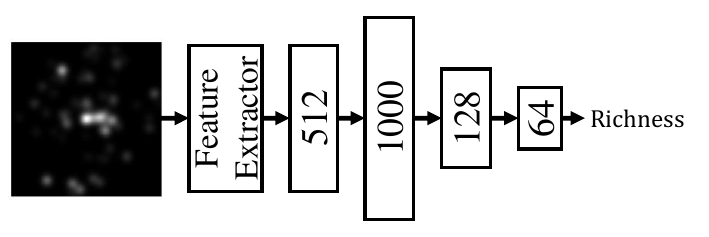}
    \caption{The full architecture of our model. The feature extractor represents ResNet-34 pre-trained model and the blocks with numbers represent the fully connected layers comprising with the corresponding number of units. The output is the richness of a galaxy cluster.}
    \label{fig:figure7}
\end{figure}

Overall, the full architecture of our model consists of two parts as shown in Figure \ref{fig:figure7}: feature extractor and four FC layers where each layer contains 512, 1000, 128, and 64 units, respectively. The former is the ResNet-34, serving as the backbone model responsible for extracting features from the images, and the four FC layers comprise a regression network. 
The choice of 512, 1000, 128, and 64 units in the FC layers is based on a combination of empirical results and the need to balance model complexity and performance. The first FC layer with 512 units reduces the dimensionality of the feature maps extracted by ResNet-34, while the subsequent layers (1000, 128, and 64 units) progressively refine the feature representation. This hierarchical reduction helps in capturing complex patterns and interactions within the data, ultimately leading to a more accurate richness estimation. 
ResNet-34 plays a crucial role in extracting information and high-order features from the SOMs of galaxy clusters. These features serve as input for the subsequent regression network. This two-step process allows us to leverage the powerful feature extraction capabilities of ResNet-34 to capture meaningful patterns and information from the SOMs, ultimately enabling accurate estimation of galaxy cluster richness.

It is noteworthy that the SOMs of galaxy clusters are 2-dimensional images with only one channel. To accommodate this, we modify the first convolutional layer of the original ResNet-34 accordingly. As illustrated in Figure \ref{fig:figure7}, data flow from left to right in the Richness Estimator. Specifically, an image with dimensions of 1 $\times$ 200 $\times$ 200 undergoes transformation into a feature vector with dimensions of 512$\times$1 through feature extraction. This feature vector is then fed into the regression network, ultimately yielding a richness prediction.

\begin{figure}
    \includegraphics[width=\columnwidth]{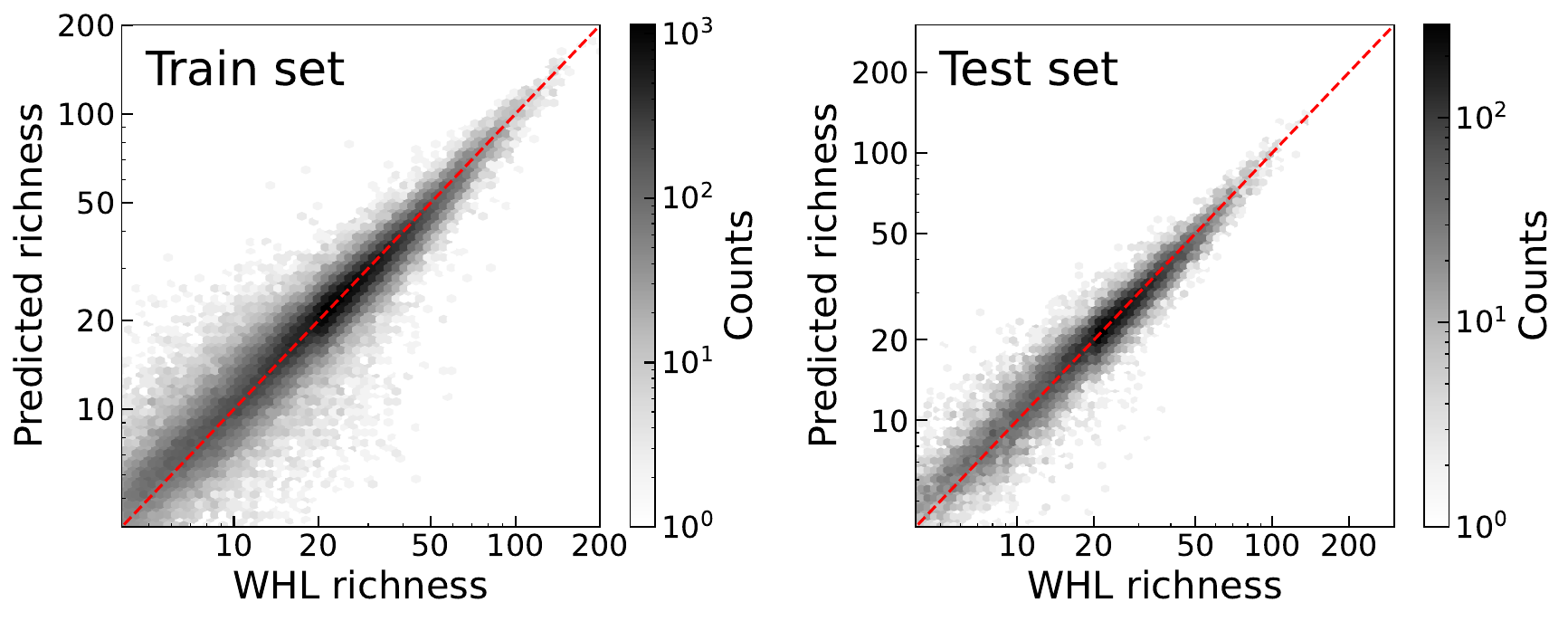}
    \caption{The richness comparison between our predicted richness and WHL richness in the training and test sets.}
    \label{fig:figure8}
\end{figure}

The loss function we use is the mean square error and the SGD optimizer\footnote{\href{https://pytorch.org/docs/stable/generated/torch.optim.SGD.html}{https://pytorch.org/docs/stable/generated/torch.optim.SGD.html}} is used to update the parameters of the model. We adopt a dynamic learning rate scheduler with an initial learning rate of 0.001. To improve the performance of the model and reduce the training epochs, the weights we load to the ResNet-34 are pretrained on the ImageNet~\citep{Deng2009}.
The performance of Richness Estimator is evaluated by employing a test set mentioned in Section \ref{sec:Smoothed optical map for training cluster richness}, and enhanced by minimizing the loss function i.e. the difference between the ground truth and the predictions. Eventually, the model reaches a convergence as the loss of test set is no longer reduced. The results of model training are depicted in the Figure \ref{fig:figure8}. The left and right panels respectively show the comparison of richness estimates for galaxy clusters in the training and test sets. 
Our richness estimates demonstrate a strong consistency with the WHL richness for clusters with WHL richness greater than 10, exhibiting biases of 0.035 dex in the training set and 0.036 dex in the test set, respectively. The associated dispersions are 0.064 dex and 0.063 dex, respectively. 
The slight difference in dispersion between the training and test sets indicates that the model has not overfit and possesses good extrapolation performance.

Different from object detection directly from color images to identify galaxy clusters, our model can provide more information about the clusters, such as their BCGs and the richness estimations, which is a great supplement to study the mass and other features of galaxy clusters. Our method demonstrates strong performance in galaxy cluster detection and can complement existing galaxy cluster detection methods. It holds promising potential for application in upcoming large-scale surveys.

\section{RESULTS AND DISCUSSION}
\label{sec:resultsANDdiscussion}

\subsection{Test data for COSMIC algorithm}

\begin{figure}
    \includegraphics[width=\columnwidth]{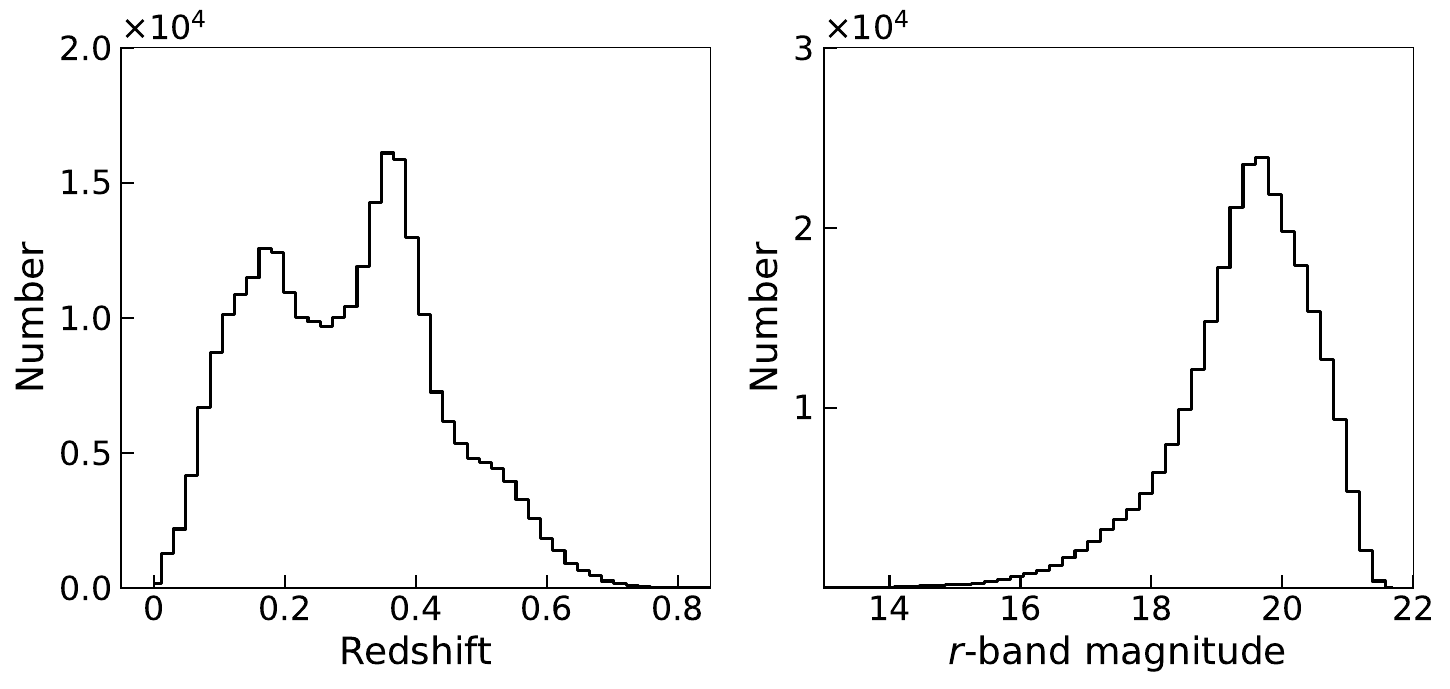}
    \caption{The redshift and $r$-band magnitude distributions of the test data.}
    \label{fig:figure9}
\end{figure}

To evaluate the performance of COSMIC algorithm, we obtain a test sample of galaxies from the SDSS database in the northern Galactic cap with R.A. from 150 to 160 degrees and Dec. from 0 to 20 degrees. This catalog serves as a vital independent test dataset, allowing us to assess the performance and robustness of our model beyond training and test. 
For this galaxy catalog, the spectroscopic redshifts are used if they are available. We make a pre-processing that the galaxies with a photometric redshift error more than $0.06(1+z)$ are excluded. Additionally, the galaxies with a detection signal-to-noise less than 10 are discarded to ensure the accuracy of the photometry.
We also exclude some outliers that may have exposure problems and some stars that are misclassified as galaxies by considering the relationship between $M_{\rm psf}^{i} - M_{\rm model}^{i}$ and $i$-band magnitude. The test sample includes 271029 galaxies.
Figure \ref{fig:figure9} shows the redshift and $r$-band magnitude distribution of galaxies of this catalog. The figure shows that the redshift of the test data exhibit a peak at around 0.4, a peak of the $r$-band magnitude of 19.

\subsection{Galaxy clusters identified from test data} \label{sec:TestInGCAP}

We apply COSMIC algorithm to the test data. In the first step for the BCGs, it is worth noting that we take into consideration the impact of thresholds when applying our model. The threshold represents a specific probability by which the model classifies an object as a BCG. A lower threshold allows the model to classify more galaxies as BCGs, but it may result in more misclassifications. Conversely, a higher threshold sets stricter criteria but might exclude some true BCGs with less distinctive characteristics. To strike a balance and identify more potential BCGs, we opt for a relatively smaller threshold of 0.4. Although this choice may lead to more misclassifications, we believe that pseudo-BCGs would be excluded during Richness Estimation.

\begin{figure}
    \includegraphics[width=\columnwidth]{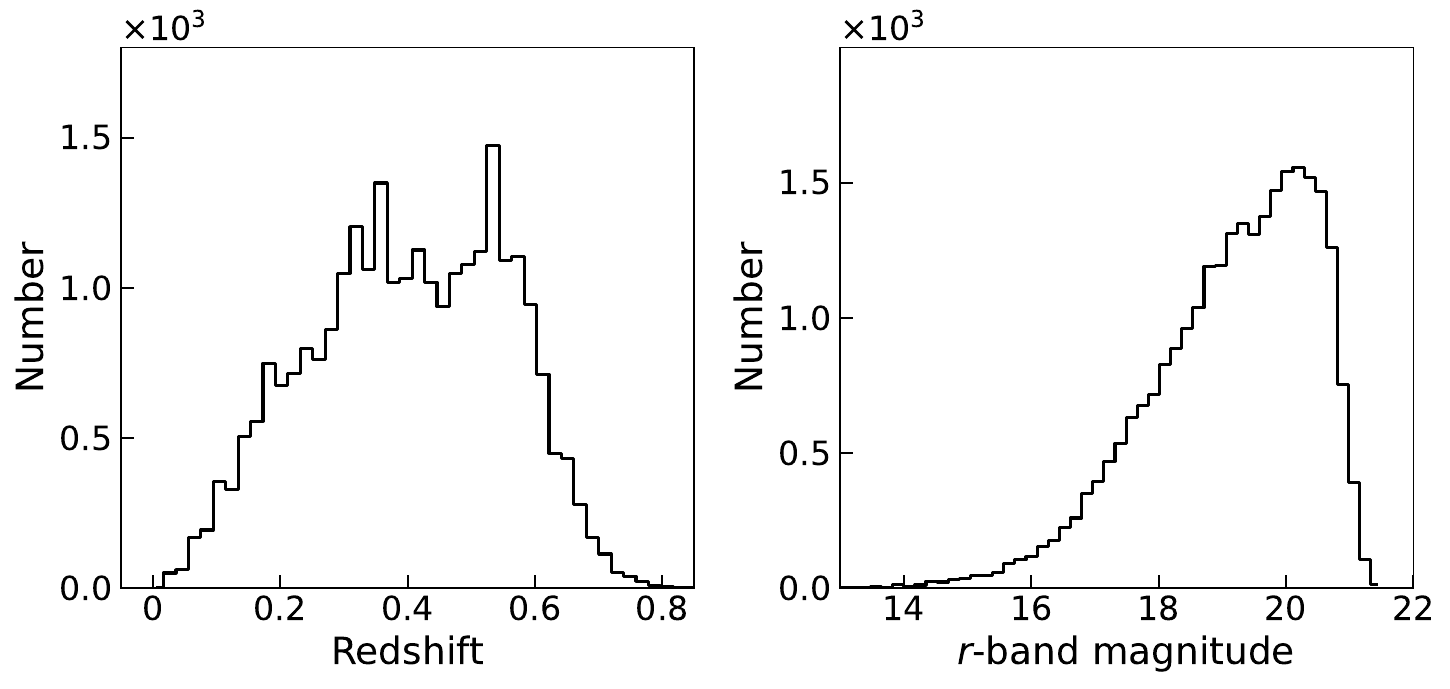}
    \caption{The redshift and $r$-band magnitude distributions of BCG candidates in the test data.}
    \label{fig:figure10}
\end{figure}

\begin{figure}
    \includegraphics[width=\columnwidth]{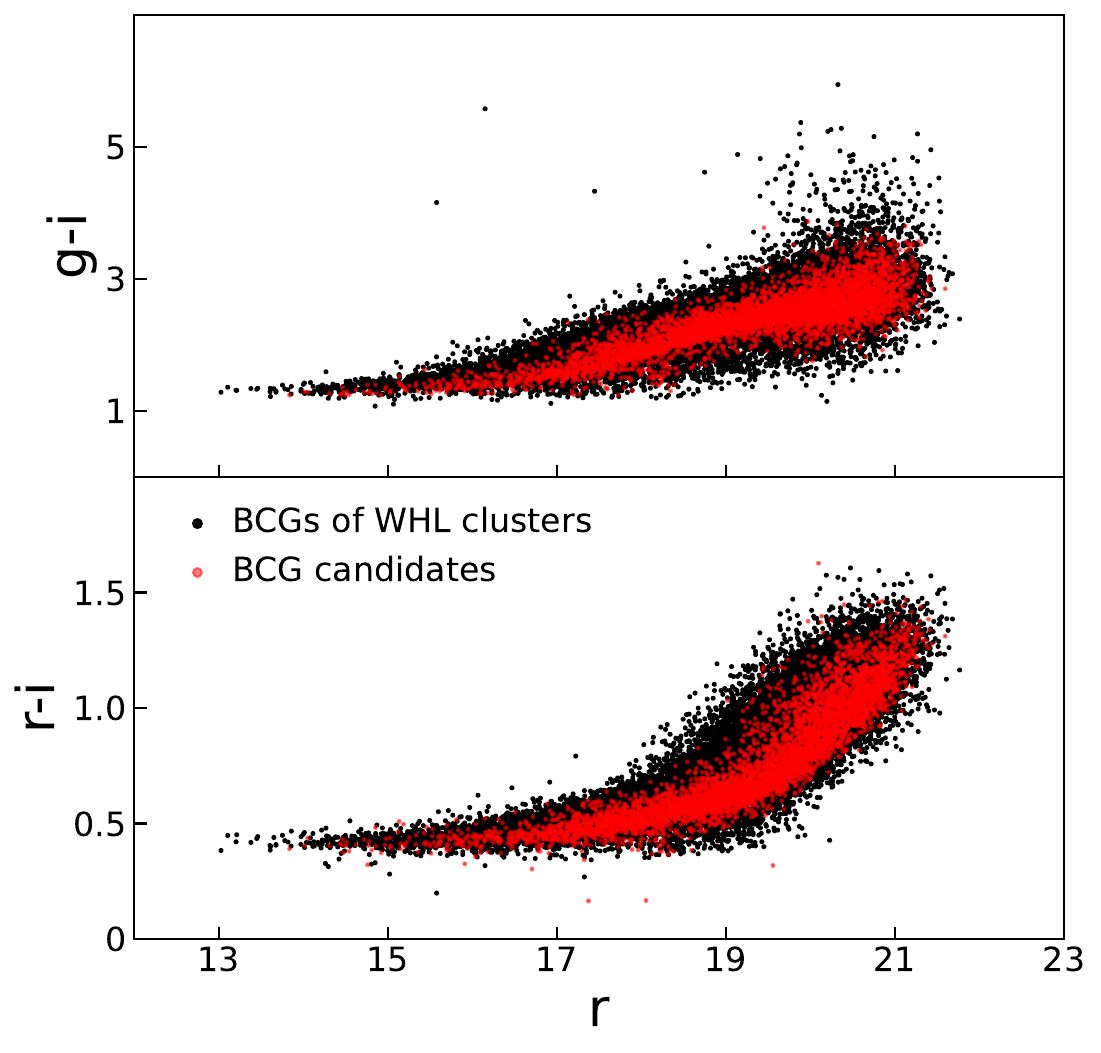}
    \caption{Color-magnitude diagram of BCG candidates compared with WHL clusters. The top panel displays the $g-i$ color, while the bottom panel shows the $r-i$ color. The horizontal axis represents the $r$-band magnitude for both BCG candidates (red dots) and the BCGs of WHL galaxy clusters (black dots), where only 20\% of them are randomly selected for plotting.}
    \label{fig:figure11}
\end{figure}

As a result, we obtain 26,717 BCG candidates from the test data. The redshift and $r$-band magnitude distributions of BCG candidates are shown in Figure \ref{fig:figure10}. We can see that the BCG candidates are distributed within the redshift range of $0.05<z<0.8$, with the majority having a redshift below 0.6. We examine them by comparing their color-magnitude diagram with that of the WHL BCGs in Figure \ref{fig:figure11}. The figure shows that our BCG candidates exhibit a good consistency with the BCGs of WHL clusters.

Subsequently, for each BCG candidate at $z$, we it get member galaxy candidates within the redshift slice of $z\pm \Delta z$ and generate SOMs, and then derive richness estimations using the Richness Estimator. According to the scaling relation between $M_{\mathrm{500}}$ (or $r_{\mathrm{500}}$) and richness in \citet{wen_calibration_2015}, we can obtain the $M_{\mathrm{500}}$ and $r_{\mathrm{500}}$ for each candidate. 
We also count the number of member galaxy candidates $N_\mathrm{gal}$ within $r_{\mathrm{500}}$ around BCG candidates. 

Considering the potential presence of multiple BCG candidates within a single galaxy cluster, we conduct a self-cross-matching of BCG candidates. This crucial step involves selecting the BCG with the highest richness from BCG candidates in close proximity as the final BCG for the cluster. The remaining candidates are considered as member galaxies within that galaxy cluster. The self-cross-matching range is confined to a projected distance of $1.5\,r_{\mathrm{500}}$, with a redshift difference less than $0.06(1+z)$. 

Finally, we identify 3653 galaxy clusters with a richness $\geq$ 10 and $N_\mathrm{gal} \geq 6$ (including the BCG), of which 1539 ($\sim$ 40\%) are newly identified from the SDSS data. The redshifts of clusters are estimated by the mean photometric redshifts of the member galaxies, or spectroscopic redshifts if available.
The parameters of a cluster are listed in Table \ref{tab:table1}, including the cluster ID, coordinates of the BCG, cluster redshift, the $r$-band magnitude of the BCG, as well as the cluster richness, $r_{500}$, $M_{\mathrm{500}}$ and the number of member galaxy candidates within $r_{500}$. 

\begin{figure}
    \includegraphics[width=\columnwidth]{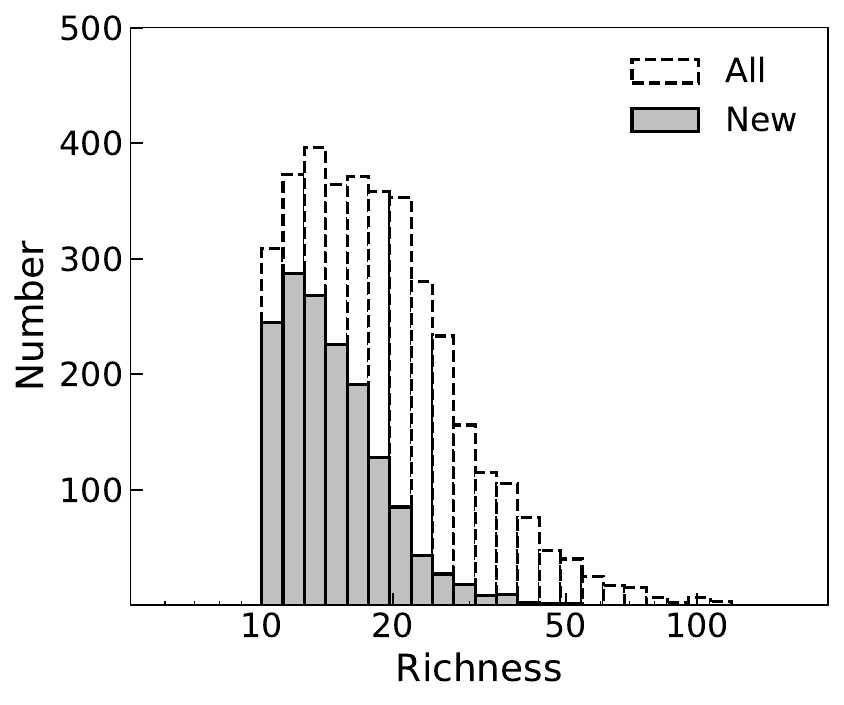}
    \caption{The richness distributions of all clusters (dashed line) and new ones (solid line).}
    \label{fig:figure12}
\end{figure}

\begin{figure*}
    \includegraphics[width=\textwidth]{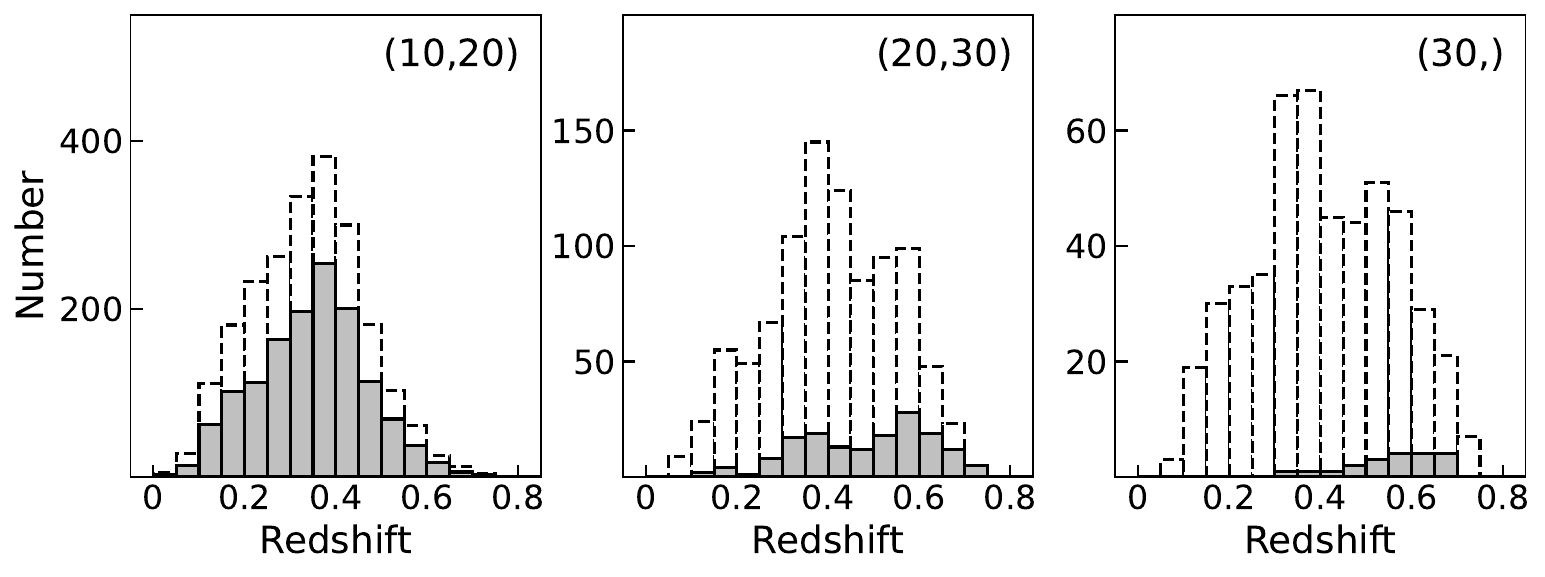}
    \caption{The richness distributions of all clusters (dashed line) and new ones (solid line) under three richness bins. }
    \label{fig:figure13}
\end{figure*}

The richness distribution of the identified clusters is shown in Figure \ref{fig:figure12}. 
The proportion of newly identified clusters is higher at lower richness intervals. This is due to the fact that the completeness of rich galaxy clusters in known catalogs is typically higher than that of poor galaxy clusters.
More specifically, when categorized into different richness intervals, the redshift distribution of new identified clusters also exhibits similar characteristics, as shown in Figure \ref{fig:figure13}. At lower redshifts, where known galaxy cluster catalogs are more complete, the newly identified clusters contain fewer relatively rich galaxy clusters, with most of them being less rich. In contrast, at higher redshifts, where known galaxy cluster catalogs are less complete, more rich clusters are identified. Overall, Figure \ref{fig:figure12} and \ref{fig:figure13} show our method detects more low-mass and high-redshift galaxy clusters, demonstrating the potential and practical applications of our approach in these aspects.

Besides the aforementioned galaxy clusters, the remaining 85\% of the BCG candidates found in the first step exhibit very low richness. These candidates could serve as potential targets for further investigation as BCG-like galaxies and lower mass galaxy clusters.

\begin{table*}
\caption{The clusters of galaxies identified from the test data.}
\begin{center}
\setlength{\tabcolsep}{1mm}
\begin{tabular}{lccccccccl}
\hline
\mc{1}{c}{Cluster ID}&\mc{1}{c}{R.A.} & \mc{1}{c}{Dec.} & \mc{1}{c}{$z_{\rm cl}$} & \mc{1}{c}{$r_{\rm BCG}$} & \mc{1}{c}{Richness} &
\mc{1}{c}{$r_{500}$} & \mc{1}{c}{$M_{500}$} &\mc{1}{c}{$N_{\rm gal}$} & \mc{1}{c}{Other catalogs} \\
\mc{1}{c}{(1)} & \mc{1}{c}{(2)} & \mc{1}{c}{(3)} & \mc{1}{c}{(4)} & \mc{1}{c}{(5)} & 
\mc{1}{c}{(6)} & \mc{1}{c}{(7)} & \mc{1}{c}{(8)} & \mc{1}{c}{(9)} & \mc{1}{c}{(10)} \\
\hline
1 & 150.0029 & 9.3456 & 0.317 & 17.02 & 40.09 & 0.82 & 2.14 & 22 & redMaPPer, WHL, Yang+21, Zou+21, CluMPR \\
2 & 150.0057 & 7.5070 & 0.359 & 18.63 & 11.38 & 0.51 & 0.54 & 7 &  \\
3 & 150.0145 & 12.2346 & 0.328 & 18.84 & 10.74 & 0.50 & 0.50 & 8 &  \\
4 & 150.0222 & 1.3500 & 0.220 & 17.25 & 17.38 & 0.62 & 0.85 & 15 & WHL, Yang+21, Zou+21, CluMPR \\
5 & 150.0284 & 6.1217 & 0.414 & 18.86 & 16.76 & 0.57 & 0.82 & 8 & GMBCG, redMaPPer, Yang+21, CluMPR \\
6 & 150.0289 & 10.6355 & 0.248 & 17.77 & 11.32 & 0.53 & 0.53 & 7 &  \\
7 & 150.0303 & 5.3786 & 0.462 & 20.28 & 21.20 & 0.60 & 1.06 & 10 & GMBCG, WHL \\
8 & 150.0309 & 6.1264 & 0.567 & 20.89 & 25.55 & 0.62 & 1.31 & 7 & WHL, Yang+21, CluMPR \\
9 & 150.0316 & 0.2980 & 0.571 & 19.88 & 33.48 & 0.69 & 1.76 & 9 & GMBCG, WHL, Yang+21, Zou+21 \\
10 & 150.0336 & 10.0234 & 0.389 & 19.13 & 35.76 & 0.76 & 1.89 & 19 & GMBCG, redMaPPer, WHL, Yang+21, Zou+21, CluMPR \\
\hline
\end{tabular}
\end{center}
\tablecomments{
Column 1: Cluster ID;
Columns 2 and 3: Right Ascension (R.A. J2000) and Declination (Dec. J2000) of cluster BCG (in degree);
Column 4: cluster redshift $z_{\rm cl}$; 
Columns 5: BCG magnitudes (AB system) in $r$-band;
Column 6: cluster richness;
Column 7: cluster radius, $r_{500}$, in Mpc; 
Column 8: derived cluster mass, in units of $10^{14}~M_{\odot}$;
Column 9: number of member galaxy candidates within $r_{500}$;
Column 10: Reference notes for previously known clusters: redMaPPer~\citep{Rykoff2016}, WHL \citep{wen_calibration_2015}, Yang+21 \citep{Yang2021}, Zou+21 \citep{Zou2021}, CluMPR \citep{YantovskiBarth2023}, GMBCG \citep{hao_gmbcg_2010}.\\
(This table is available in its entirety in a machine-readable form.)
}
\label{tab:table1}
\end{table*}

\subsection{Cross-match with previous SDSS cluster catalogs} \label{sec:cross match}

To verify whether the known galaxy clusters can be detected by COSMIC algorithm, 
we compare our clusters with the clusters in four SDSS cluster catalogs, including the MaxBCG~\citep{koester_maxbcg_2007}, GMBCG~\citep{hao_gmbcg_2010}, redMaPPer~\citep{Rykoff2016} and WHL~\citep{wen_catalog_2012,wen_calibration_2015}. 
We perform the cross-matching within a redshift difference of $0.05(1+z)$ within the matching radius of 2 arcsec and a projected separation of 1.5$r_{\mathrm{500}}$, respectively. The choice of $0.05(1+z)$ for the redshift difference in cross-matching is sufficiently large to include the same galaxy cluster across different catalogs. This threshold corresponds to roughly 3 times the typical photometric redshift uncertainties of clusters observed in various catalogs. For instance, the redshift error $\sigma_z$ is  $< 0.02$ for the redMaPPer at $z<0.5$, $< 0.01$ for the MaxBCG, $< 0.015$ for the GMBCG, and $< 0.018$ for the WHL. Given that $0.05(1+z)$ is approximately $2.5-5 \sigma_z$ for these catalogs, it can serve as a reasonable threshold to ensure proper cross-matching of clusters from different catalogs.

\begin{figure*}
    \includegraphics[width=\textwidth]{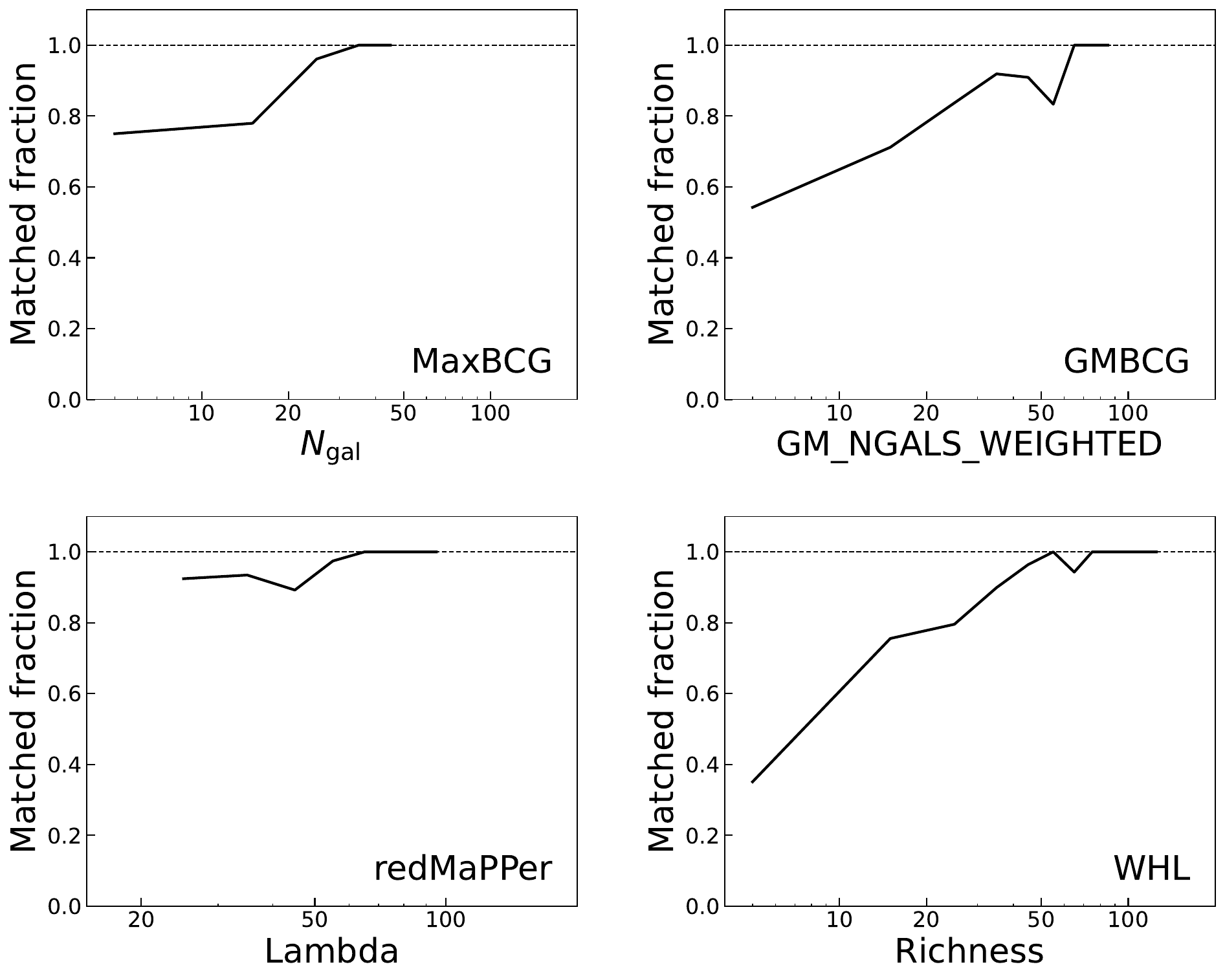}
    \caption{The matched fraction across different richness intervals of four SDSS cluster catalogs.}
    \label{fig:figure14}
\end{figure*}

As shown in Table \ref{tab:table2}, COSMIC algorithm recovers about 46\% BCGs of the GMBCG
catalog in the same sky region, and more than 57\% BCGs of the maxBCG, WHL and redMaPPer
catalogs.
Noticed that a cluster may contain multiple massive BCG-like galaxies with similar luminosities. 
These algorithms determine the cluster centers on different BCG-like galaxies. Adopting a larger matching radius is more reasonable. We find that our algorithm finds about 67\% GMBCG clusters within a radius of 1.5$r_{\mathrm{500}}$, and 81\% maxBCG, 78\% WHL and 93\% redMaPPer clusters, respectively. 
Figure \ref{fig:figure14} shows the matched fraction by our catalog as a function of 
the cluster richness for clusters in previous catalogs. 
The matched fraction increases with richness and reaches near 100\% for a richness $> 60$. The matched fraction is more than 50\% even for the clusters with a low richness about 10. 
These results indicate that COSMIC algorithm effectively finds previous known galaxy clusters.

\begin{table*}
\caption{The results of cross-matching with SDSS cluster catalogs.}
\begin{center}
\setlength{\tabcolsep}{1mm}
\begin{tabular}{cccccc}
    \hline
    \mc{1}{c}{catalog} & \mc{1}{c}{N} & \mc{1}{c}{N$_{\mathrm{2 arcsec}}$} & 
    \mc{1}{c}{C$_{\mathrm{2 arcsec}}$} & \mc{1}{c}{N$_{1.5r_{\mathrm{500}}}$} & 
    \mc{1}{c}{C$_{1.5r_{\mathrm{500}}}$} \\
    \mc{1}{c}{(1)} & \mc{1}{c}{(2)} & \mc{1}{c}{(3)} & \mc{1}{c}{(4)} & 
    \mc{1}{c}{(5)} & \mc{1}{c}{(6)} \\ 
    \hline
    MaxBCG & 320 & 183 & 57.19\% & 260 & 81.25\% \\
    GMBCG & 1121 & 519 & 46.30\% & 746 & 66.55\% \\
    redMaPPer & 442 & 293 & 66.29\% & 412 & 93.21\% \\
    WHL15 & 2209 & 1382 & 62.56\% & 1729 & 78.27\% \\
    \hline
    \end{tabular}
\end{center}
\tablecomments{
    Column (1): catalog name; 
    Column (2): total number of galaxy clusters in test data; 
    Column (3) and (5): number of cross-matched clusters; 
    Column (4) and (6): matching rate of cross-matched clusters at a range of 2 arcsec and 1.5$r_{\mathrm{500}}$.
}
\label{tab:table2}
\end{table*}

Despite successfully recovering most of the galaxy clusters in these catalogs, there still exist a few omissions. There are three main reasons that failed to recover. 
First, only about 25\% of missed clusters are not detected by our method, where most of them show low BCG prediction probabilities and are below the threshold. 
Secondly, some clusters contain multiple BCG-like galaxies, and due to variations in BCG selection methods, our identified BCGs may differ from those in other catalogs, particularly for merging clusters. However, due to uncertainties in photometric redshift measurements, the redshift discrepancies of the BCGs may exceed the matching threshold, resulting in non-matching cases.
Thirdly, the remaining clusters contain BCG-like galaxies, but their richness and the total number of member galaxy candidates fall below our thresholds, particularly for poor and high-redshift clusters. 
Indeed, with accurate redshift information for galaxies, it should be feasible to match most of rich clusters by our method. This further underscores the effectiveness of our method in accurately identifying galaxy clusters.

\begin{figure*}
    \includegraphics[width=\textwidth]{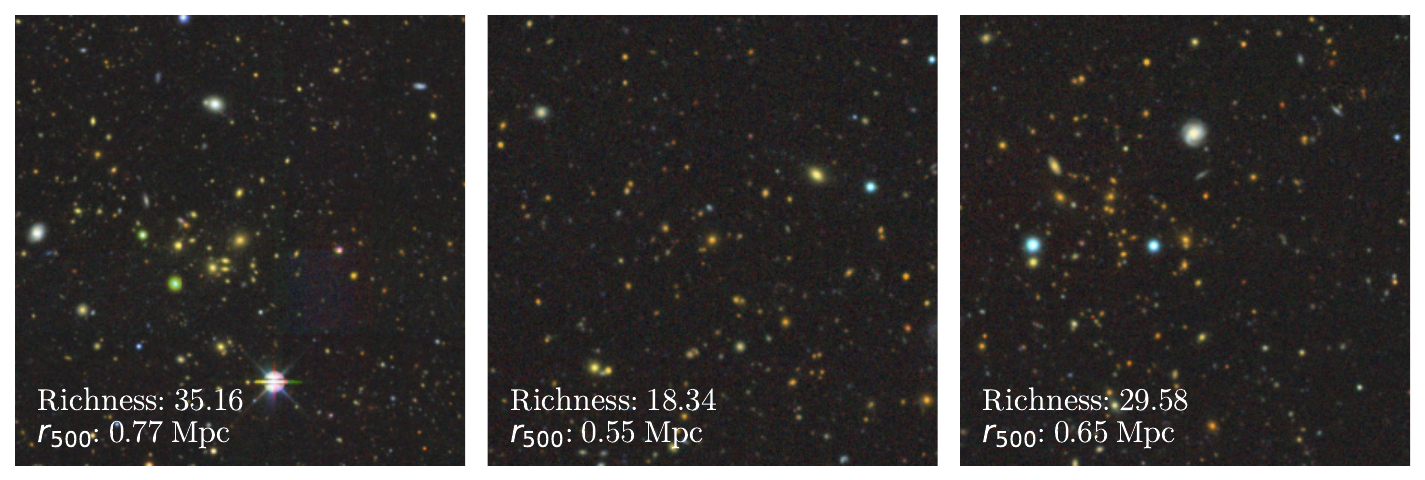}
    \caption{Color images of three newly identified clusters from DESI Legacy Surveys. The physical radius of the images is equal to their $r_{\mathrm{500}}$.}
    \label{fig:figure15}
\end{figure*}

In addition to finding known clusters in the above four SDSS catalogs, our method is able to find clusters that are not included in the previous SDSS catalogs, accounting for about 40\% of the total.
As suggested in Figure \ref{fig:figure13}, our method is promising to find more poor and hgih-redshift clusters. Figure \ref{fig:figure15} shows colour images from DESI Legacy Surveys\footnote{\href{https://www.legacysurvey.org/viewer}{https://www.legacysurvey.org/viewer}} of three newly identified clusters. 

\subsection{Richness comparison with previous optical cluster catalogs} \label{sec:Richness comparison to the optical cluster catalogs}

\begin{figure}
    \includegraphics[width=\columnwidth]{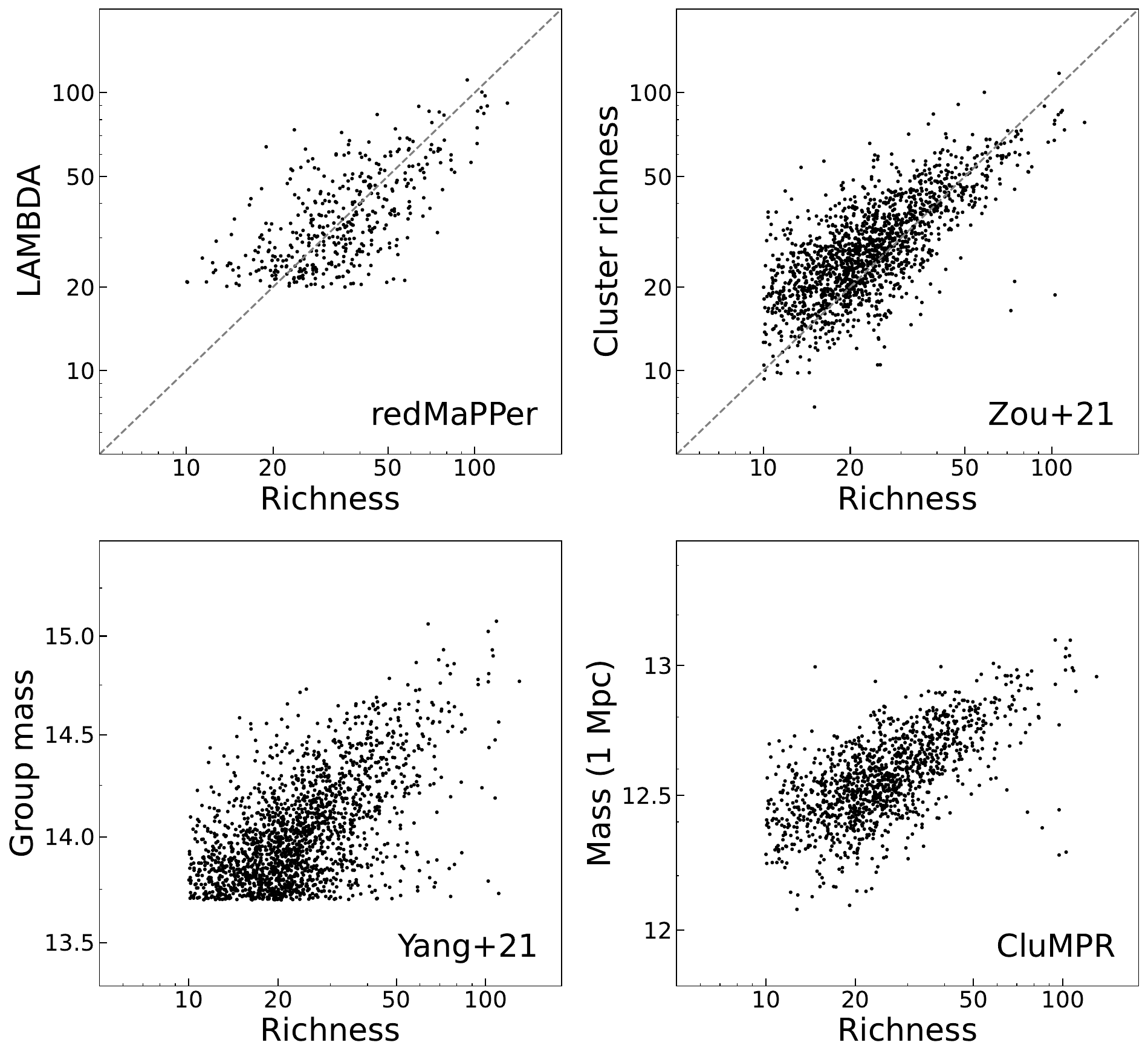}
    \caption{Richness comparison with optical cluster catalogs. The top two panels show the comparison to cluster richness in redMaPPer and Zou+21 catalogs. The bottom two panels show the comparison to the group mass in Yang+21 catalog and the log stellar mass of the cluster within a 1-Mpc physical radius which corrected for incompleteness and galaxy background in CluMPR catalog. The dashed line represents $y = x$.}
    \label{fig:figure16}
\end{figure}

We compare our estimated richness with those in one SDSS cluster catalog and three cluster catalogs from DESI Legacy Surveys. Based on the SDSS data, the redMaPPer richness is defined as to be the sum over the members with an estimated membership probability. Based on the DESI Legacy Surveys data,
Yang+21 \citep{Yang2021} estimated the halo masses of the clusters by
the total luminosities of cluster member galaxies for the clusters by an extended halo-based group finder, Zou+21 \citep{Zou2021} defined the richness to be the total
luminosity of member galaxy candidates for the CFSFDP clusters. For the CluMPR \citep{YantovskiBarth2023} clusters, we use the total stellar mass of clusters within 1 Mpc for comparison.
In Figure \ref{fig:figure16}, we conduct a comparative analysis between our richness estimates and those provided by radMaPPer, Zou+21, Yang+21 and CluMPR. It is clear that our richness estimates demonstrate a robust correlation with those from other galaxy cluster catalogs. The scatters of correlation for radMaPPer, Zou+21, Yang+21 and CluMPR are 0.095, 0.100, 0.176 and 0.191 dex, respectively.
This results illustrate the reliability of our method for estimating richness.

\subsection{Compare to the X-ray cluster catalogs}

Recently, the German eROSITA Consortium (eROSITA-DE) has publicly released the initial six months of data from the SRG/eROSITA all-sky survey (eRASS1), which provides the largest X-ray cluster catalog \citep{Kluge2024}. The eROSITA catalog comprises optical characteristics of the eRASS1 clusters and groups, specifically the extended X-ray sources from \citet{Bulbul2024}, for which optical properties such as redshift and richness are determined using optical and near-infrared data from the DESI Legacy Surveys. This X-ray cluster catalog provides another way for evaluating our method.

\begin{figure}
    \includegraphics[width=\columnwidth]{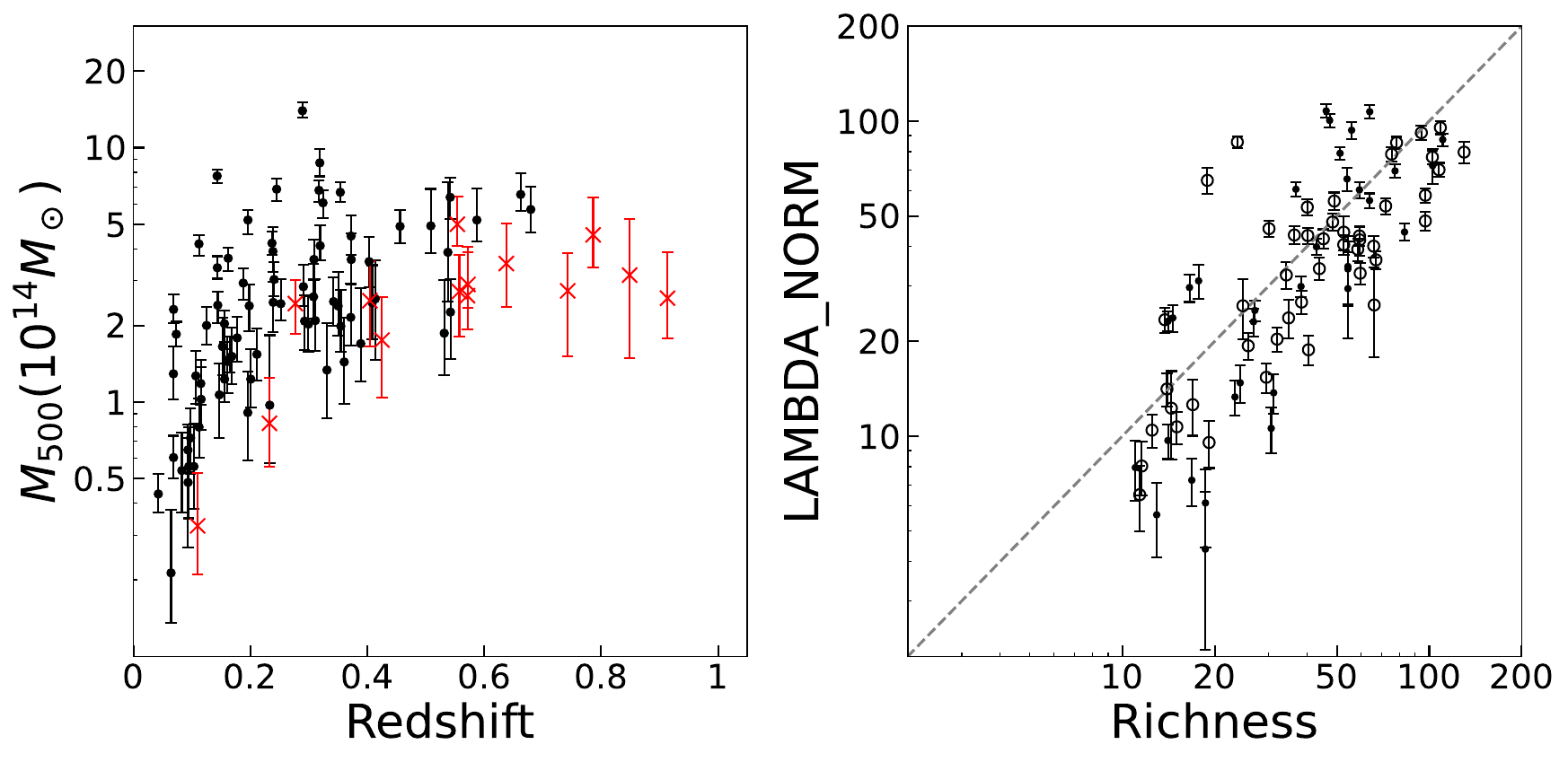}
    \caption{The left panel shows the matched (black dots) and un-matched (red crosses) eROSITA clusters in the mass-redshift diagram. The right panel shows the correlations between our estimated richness and the richness (LAMBDA\_NORM) in eROSITA catalog. The circles and dots represent the matched clusters using different matching separations for 2 arcsec and 1.5$r_{\mathrm{500}}$, respectively. The dashed line represents $y = x$.}
    \label{fig:figure17}
\end{figure}

We get 103 eROSITA X-ray clusters located in the test sky region. COSMIC algorithm finds 80 eROSITA 
clusters within a redshift difference of $0.05(1+z)$ and a projected separation of 1.5\,$r_{\mathrm{500}}$.
The left panel of Figure \ref{fig:figure17} shows the matched and unmatched eROSITA clusters in the 
mass-redshift diagram. The panel only displays clusters for which $M_{\mathrm{500}}$ has been measured.
Noted that the eROSITA and DESI Legacy Surveys data are deeper than the SDSS for clusters. The unmatched clusters mostly distribute in $z>0.4$ or low-mass region. 

We check the clusters failed to matched in eROSITA catalog.  
Overall, there are 23 missed eROSITA clusters, with 16 of high redshift BCGs actually not found in the SDSS galaxy catalog.
We find that there is an eROSITA galaxy cluster that fails to meet the pre-processing error criteria. Four unmatched clusters have richness that are too low to satisfy the threshold. For the remaining three clusters, their BCGs are misclassified due to either low richness or significant uncertainties in their photometric redshifts.
In summary, the main reasons for the incompleteness are attributed to several factors:
First, the eROSITA covers a broader redshift range, and some high-redshift clusters fall outside the detection limits of the SDSS. Secondly, the purity of eROSITA catalog is only 86\%, meaning some identified clusters are not real. When checked with DESI Legacy Surveys, some of these clusters do not appear to have the galaxy overdensity characteristic of true clusters. Finally, we accurately identify high-mass clusters, while the detection of low-mass clusters is more uncertain due to their lower richness and the limitations of our data.

In right panel of Figure \ref{fig:figure17}, we show the correlation between our estimated richness and the richness (LAMBDA\_NORM)
in the eROSITA catalog, where the circles and dots represent the matched clusters using different matching separations for 2 arcsec and 1.5\,$r_{\mathrm{500}}$, respectively. The richness comparison shows the scatters of 0.126 and 0.134 dex for 2 arcsec and 1.5$r_{\mathrm{500}}$, respectively. 
Again, the correlation confirms the reliability of our richness. 

\begin{figure}
    \includegraphics[width=\columnwidth]{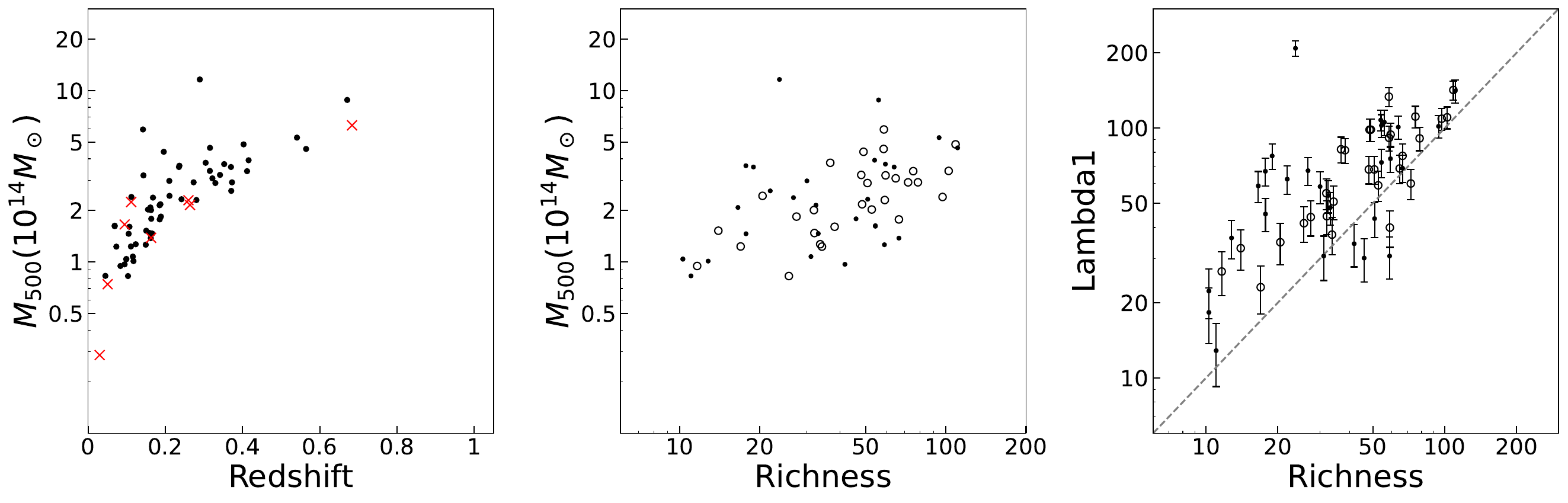}
    \caption{The left panel shows the matched (black dots) and un-matched (red crosses) RASS-MCMF clusters in the mass-redshift diagram. The right two panels show the correlations of our estimated richness with both $M_{\mathrm{500}}$ and the richness (Lambda1) in RASS-MCMF catalog. The circles and dots represent the matched clusters using different matching separations for 2 arcsec and 1.5$r_{\mathrm{500}}$, respectively. The dashed line represents $y = x$.}
    \label{fig:figure18}
\end{figure}

We also compare our identified clusters with the RASS-MCMF \citep{Klein2023} and MCXC-II \citep{Sadibekova2024} X-ray cluster catalogs. These are two comprehensive compilations of X-ray selected galaxy clusters. The RASS-MCMF catalog includes 8449 clusters detected by the ROSAT All-Sky Survey and confirmed using the MCMF algorithm, while the MCXC-II catalog aggregates 2221 clusters from various ROSAT-based and serendipitous surveys, providing standardized measurements of X-ray luminosity and mass. Both catalogs offer high-purity samples of X-ray selected galaxy clusters.
There are 65 and 12 clusters in the RASS-MCMF and MCXC-II within the test sky region. COSMIC algorithm identifies 87\% and 83\% of the clusters from the two catalogs within a redshift difference of $0.05(1+z)$ and a projected separation of 1.5\,$r_{\mathrm{500}}$. The clusters missed in MCXC-II, as well as three quarters of those missed in RASS-MCMF, lie beyond the SDSS photometric observation.
Here, we present only the comparison to the RASS-MCMF catalog, as shown in Figure \ref{fig:figure18}. Similar to Figure \ref{fig:figure17}, the left panel displays the matched clusters (black dots) and unmatched clusters (red crosses) in the mass-redshift diagram. We observe that the unmatched clusters are primarily located in the low-mass or high-redshift regions.
The middle and right panels show the correlations between our estimated richness and both the mass ($M_{500}$) and the richness from RASS-MCMF. In these panels, circles and dots represent the matched clusters using different matching separations of 2 arcsec and 1.5$r_{\mathrm{500}}$, respectively.
As seen in the figure, aside from one outlier, our estimated richness shows a strong correlation with both the $M_{500}$ and richness from RASS-MCMF. Upon investigating the outlier (RASS-CL J102340+0411.2), we find that the BCG is not correctly classified in our BCG Classification, with only a 10\% probability assigned, leading its SOM to deviate significantly from the cluster center and causing an underestimation of richness. The color image of the BCG from the DESI Legacy Surveys shows an unusual blue color, and notably, the optical center identified by RASS-MCMF is also offset from the BCG.
After excluding this outlier, the dispersion of our estimated richness compared to the mass and richness from RASS-MCMF is 0.236 and 0.147 dex using a matching separation of 1.5$r_{\mathrm{500}}$, and 0.185 and 0.108 dex using a separation of 2 arcsec, respectively. This further confirms the reliability of our richness estimates.

\section{SUMMARY AND PERSPECTIVE} \label{sec:summaryANDperspective}

We present a new algorithm named COSMIC to detect galaxy clusters from optical photometric data with machine learning by two steps. First, we use XGBoost to classify a complete sample of BCG-like galaxies. Then, we use ResNet34 to estimate richness of cluster candidates centering on the BCG-like galaxies.

COSMIC is applied to the photometric data from the SDSS. Within a test sky region of $\sim$200 deg$^2$, we detect 3653 galaxy clusters. Our cluster catalog contains most of rich clusters in the previous SDSS cluster catalogs, and X-ray clusters in the eROSITA X-ray cluster catalog.
Our estimated richness demonstrates a strong correlation with the richness and mass of clusters in optical catalogs, and exhibits a small dispersion when compared to cluster richness and mass in the eROSITA and RASS-MCMF X-ray cluster catalogs.
Overall, our method is capable of detecting the majority of known galaxy clusters and is also effective in identifying more low-mass or high-redshift clusters.

In conclusion, it has been demonstrated that machine learning techniques are a promising alternative to conventional methods for identifying galaxy clusters. As large-scale surveys, such as Euclid and 
Chinese Space Station Telescope, accompanied by an unprecedented increase in observational data, we aim to fully leverage these datasets to optimize our models and achieve superior performance. We anticipate that the forthcoming cluster catalogs generated by our approach, either alone or in conjunction with complementary detection techniques, will be more comprehensive and thus conducive to further understand our universe.


\section*{acknowledgments}
We thank the referee for helpful comments.
J.-Q.X. is supported by the National Natural Science Foundation of China, under grant Nos. 12473004 and 12021003, the National Key R\&D Program of China, No. 2020YFC2201603, the China Manned Space Program through its Space Application System, and the Fundamental Research Funds for the Central Universities. Z.-L.W. is supported by the National Natural Science Foundation of China, under grant No. 12073036 and the science research grants from the China Manned Space Project with Numbers CMS-CSST-2021-A01.

Funding for the Sloan Digital Sky Survey IV has been provided by the 
Alfred P. Sloan Foundation, the U.S. Department of Energy Office of 
Science, and the Participating Institutions. 

SDSS-IV acknowledges support and resources from the Center for High 
Performance Computing  at the University of Utah. The SDSS website is www.sdss.org.

SDSS-IV is managed by the Astrophysical Research Consortium 
for the Participating Institutions of the SDSS Collaboration including 
the Brazilian Participation Group, the Carnegie Institution for Science, 
Carnegie Mellon University, Center for Astrophysics | Harvard \& 
Smithsonian, the Chilean Participation Group, the French Participation Group, Instituto de Astrof\'isica de Canarias, The Johns Hopkins 
University, Kavli Institute for the Physics and Mathematics of the 
Universe (IPMU) / University of Tokyo, the Korean Participation Group, 
Lawrence Berkeley National Laboratory, Leibniz Institut f\"ur Astrophysik 
Potsdam (AIP),  Max-Planck-Institut f\"ur Astronomie (MPIA Heidelberg), 
Max-Planck-Institut f\"ur Astrophysik (MPA Garching), 
Max-Planck-Institut f\"ur Extraterrestrische Physik (MPE), National Astronomical Observatories of China, New Mexico State University, New York University, University of Notre Dame, Observat\'ario 
Nacional / MCTI, The Ohio State University, Pennsylvania State University, Shanghai Astronomical Observatory, United Kingdom Participation Group, Universidad Nacional Aut\'onoma 
de M\'exico, University of Arizona, University of Colorado Boulder, 
University of Oxford, University of Portsmouth, University of Utah, 
University of Virginia, University of Washington, University of Wisconsin, Vanderbilt University, and Yale University.


\bibliography{ref}{}
\bibliographystyle{aasjournal}



\end{document}